\documentclass[twocolumn,pre,showpacs,amsmath,amssymb,nobibnotes,nofootinbib,superscriptaddress,noshowpacs]{revtex4}
\usepackage{natbib}
\usepackage{amsmath}
\usepackage{dcolumn}
\usepackage{wrapfig}
\usepackage{graphicx}
\usepackage{color}
\usepackage{graphics}
\usepackage{epsfig}
\usepackage{dsfont}
\usepackage{wasysym}
\newcommand{\cut}[1]{{}}
\newcommand{\new}[1]{\textcolor{black}{#1}}
\newcommand{\old}[1]{}

\parskip=\medskipamount

\newcommand{\be}{\begin{equation}}
\newcommand{\ee}{\end{equation}}

\begin{document}

\title{Optimal Deployment of Resources for Maximizing Impact in Spreading Processes}

\author{Andrey Y. Lokhov}
\affiliation{Center for Nonlinear Studies and Theoretical Division T-4, Los Alamos National Laboratory, Los Alamos, NM 87545, United States}
\author{David Saad}
\affiliation{The Nonlinearity and Complexity Research Group, Aston University, Birmingham B4 7ET, United Kingdom}
\begin{abstract}
The effective use of limited resources for controlling spreading processes on networks is of prime significance in diverse contexts, ranging from the identification of ``influential spreaders'' for maximizing information dissemination and targeted interventions in regulatory networks, to the development of mitigation policies for infectious diseases and financial contagion in economic systems. Solutions for these optimization tasks that are based purely on topological arguments are not fully satisfactory; in realistic settings the problem is often characterized by heterogeneous interactions and requires interventions over a finite time window via a restricted set of controllable nodes. The optimal distribution of available resources hence results from an interplay between network topology and spreading dynamics. We show how these problems can be addressed as particular instances of a universal analytical framework based on a scalable dynamic message-passing approach and demonstrate the efficacy of the method on a variety of real-world examples.
\end{abstract}
\maketitle

Spreading corresponds to omnipresent processes describing a vast number of phenomena in social, natural and technological networks~\cite{anderson1992infectious,Boccaletti2006,Rogers2010,pastor2015epidemic} whereby information, viruses and failures propagate through their edges via the interactions between individual constituents. Spreading cascades have a huge impact on the modern world, be it negative or positive. An 11 minute power grid disturbance in Arizona and California in 2011 led to cascading outages and left 2.7 million customers without power~\cite{powergrid}. As many as 579,000 people around the world could have been killed by the H1N1 influenza pandemic characterized by a rapid spreading through the global transportation networks~\cite{dawood2012estimated}. The U.S. economy losses from the 2008 financial crisis resulted from cascading bankruptcies of major financial institutions are estimated at the level of \$22 trillion~\cite{crisis}. Therefore, it is not surprising that efficient prediction and control of these undesired spreading processes are regarded as fundamental questions of paramount importance in developing policies for optimal placement of cascade-preventing devices in power grid, real-time distribution of vaccines and antidotes to mitigate epidemic spread, regulatory measures in inter-banking lending networks and other modern world problems, such as protection of critical infrastructures against cyber-attacks and computer viruses~\cite{lokhov2016detection}.

On the other hand, spreading processes can also be considered  beneficial. The ice bucket challenge campaign in social networks raised \$115 million donations to the ALS association fighting the Amyotrophic Lateral Sclerosis, in particular due to a significant involvement of celebrities acting as ``influencers''~\cite{alsa}. In the context of political campaigning, there are already winners~\cite{rutledge2013obama,epstein2015search} and losers, and this division is likely to become more pronounced and critical in the future~\cite{margetts2015political}. Winners are those who use communication and social networks effectively to set the opinions of voters or consumers, maximizing the impact of scarce resource such as activists or advertisements by applying control to the most influential groups of nodes at the right time; while losers will spend their resource sub-optimally relying on intuition and serendipity. Additional examples of domains where optimal resource allocation plays a crucial role in enhancing the effect of spreading include viral marketing campaigns~\cite{domingos2001mining}, targeted chemically-induced control of dynamic biological processes~\cite{martin2013computational}; drug discovery~\cite{csermely2013structure}; and even gaining military advantage through the propagation of disinformation~\cite{jones2015army}. All these applications share several important common properties such as restricted budget, finite-time windows for control interventions and the need for fast and scalable optimization algorithms which can be deployed in real time.

There exists a large body of work on optimal resource deployment in various spreading settings. A widely addressed formulation focuses on identifying influential spreaders, i.e. nodes that play important role in the dynamical process. Identification is often done by employing different centrality measures based on the topology of the underlying interaction network, including selection strategies based on high-degree nodes~\cite{pastor2002immunization}, neighbors of randomly selected vertices~\cite{cohen2003efficient}, betweenness centrality~\cite{holme2002attack}, random-walk~\cite{holme2004efficient}, graph-partitioning~\cite{chen2008finding}, and k-shell decomposition~\cite{kitsak2010identification}, to name a few. It is quite natural that algorithms based exclusively on topological characteristics appear to have variable performance depending on particular network instances and dynamical models used~\cite{borge2012absence, hebert2013global}. Another line of work consists in studying the NP-complete problem of network dismantling~\cite{morone2015influence,mugisha2016identifying,braunstein2016network}: the underlying reasoning is that removal of nodes breaking the giant component to small pieces is likely to prevent the global percolation of the contagion. The localization of an optimal immunization set has been addressed using a belief propagation algorithm built on top of percolation-like equations for SIR (Susceptible, Infected, Recovered) and SIS (Susceptible, Infected, Survived) models~\cite{altarelli2014containing}, based on cavity method techniques developed previously for deterministic threshold models~\cite{altarelli2013optimizing, guggiola2015minimal}. This formulation is close to the problem of finding optimal~\emph{seeds}, i.e. the smallest set of initial nodes which maximizes the spread asymptotically~\cite{domingos2001mining}. It was rigorously analyzed~\cite{kempe2003maximizing,chen2013information} for two simple diffusion models with a special submodularity property, Independent Cascade (IC) and Linear Threshold, and was shown to be NP-hard for both. A greedy algorithm based on a sampling subroutine has been explored for the IC model~\cite{du2013scalable} in the setting of finite time horizon. For other spreading models the impact maximization problem at finite time and resources has been addressed in the setting of optimal control as reported in a recent survey~\cite{nowzari2015analysis}. However, only deterministic mean-field dynamics have been considered so far; this approximation ignores the topology of the specific network considered and yields non-distributed solutions to the control problem.

All of these techniques consider the problem of static (open-loop) resource allocation, preplanned at some initial time. A less explored direction consists in developing an online policy of assigning a limited remedial budget dynamically based on real-time feedback, also known as a closed-loop control. The impact of vaccination of the largest degree nodes or of those with the largest number of infected neighbors was investigated in~\cite{borgs2010distribute,nian2010efficient}, while an alternative strategy is focused on the largest reduction in infectious edges~\cite{scaman2015greedy}. Finally, an online policy based on the resolution of the minimal maxcut problem was introduced~\cite{drakopoulos2014efficient}, where optimization is carried out with respect to the expected time to extinction of the SIS epidemic.

We introduce a general optimization framework which accommodates both dynamical and topological aspects of the problem and which allows for a broad range of objectives. The framework is principled, probabilistic, computationally efficient and incorporates the topological properties of the specific network under consideration. It facilitates the optimization of objective functions beyond the maximization or minimization of the spread, including: targeting specific nodes at specific times given a subset of accessible nodes; a limited global budget, possibly distributed over time; and an optimal dynamic vaccination strategy using the feedback from the spreading process. The problem is stated in a dynamical control setting with finite-time horizon that requires an explicit resolution of the dynamics, which is addressed via a distributed message-passing algorithm. We test the efficacy of the method on particular synthetic optimization problems as well as on a set of real-world instances.

\section{Results}

\subsection{Model}

A large number of spreading models have been suggested in the literature to describe stochastic dynamical processes in epidemiology, information and rumor propagation, and cascades in biological and infrastructure networks~\cite{Boccaletti2006,Rogers2010,pastor2015epidemic}. They all share the same common features: the nodes transition from inactive to active state due to spontaneous activation mechanism associated with the nodes themselves, or due to interactions with active neighbours through the network edges. As an illustration of our approach, we have chosen a popular stochastic spreading process known as susceptible-infected-recovered, or SIR model, which is often used to describe propagation of infectious diseases or information spreading~\cite{Boccaletti2006}. More precisely, we consider a generalized version of the discrete-time SIR model defined as follows. A node $i$
in the interaction graph $G\!=\!(V,E)$, where $V$ denotes the set of nodes, and $E$ is the set of pairwise edges, at time step $t$ can be found in either of three states $\sigma^{t}_{i}$: ``susceptible'' $\sigma^{t}_{i}\!=\!S$, ``infected'' $\sigma^{t}_{i}\!=\!I$ or ``recovered'' $\sigma^{t}_{i}\!=\!R$. At each time step, an infected (or, depending on the application domain, informed or active) individual $i$ can transmit the activation signal to one of its susceptible (respectively, uninformed or inactive) neighbors $j$ with probability $\alpha_{ij}$, associated with the edge connecting them. Independently on the interaction between nodes a node $i$ in the $S$ state can turn active, assuming state $I$, with the probability $\nu_{i}(t)$, or spontaneously become recovered (uninterested, protected) with probability $\mu_{i}(t)$ at time step $t$. The first mechanism corresponds to a node activation due to an external influence such as advertisement in the context of information spreading. In the case of the epidemic spreading the second mechanism models the effect of vaccination: once a node goes to the protected $R$ state, it becomes immune to the infection at all times. These probabilistic transmission rules at each time step $t$ can be summarized using the following schematic rules:
\begin{align}
S(i)+I(j) &\xrightarrow{\alpha_{ji}}I(i)+I(j),\label{eq:SIR_rules_StoI_interaction}
\\
S(i) \xrightarrow{\nu_{i}(t)}I(i), &\quad\quad\, S(i) \xrightarrow{\mu_{i}(t)}R(i). \label{eq:SIR_rules_StoI_spontaneous}
\end{align}
In the definition of the dynamic rules \eqref{eq:SIR_rules_StoI_interaction} and \eqref{eq:SIR_rules_StoI_spontaneous}, $\nu_{i}(t)$ and $\mu_{i}(t)$ represent control parameters we could manipulate with a certain degree of freedom defined by a particular instance of the problem. In what follows, we assume that the spreading couplings $\alpha_{ij}$ are known (or can be estimated) and are fixed in time. In some applications, $\alpha_{ij}$ may vary in time (e.g. this is true for temporal networks) or may represent a set of control parameters themselves. We outline such scenarios in the~\emph{Discussion} section; the optimization scheme presented below can be straightforwardly generalized to include the edge-related control parameters. However, for simplicity we will only present optimization involving node-related control parameters.

\begin{figure}[!th]
\centerline{\includegraphics[width=0.84\linewidth]{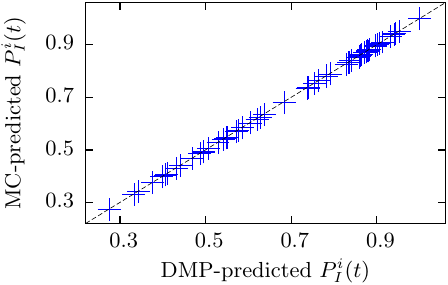}}
\caption{Performance of DMP equations for the generalized SIR model on a network of flights between major US airports. The network represents $M\!=\!383$ flight routes between the $N\!=\!61$ largest US hubs. The weights $\alpha_{ij}$ are proportional to the average number of transported passengers on each route and are distributed in the interval $[0.05,0.5]$; $\nu_{i}$ and $\mu_{i}$ are generated at random in the range $[0,0.1]$. The scatter plot represents marginal probabilities $P^{i}_{I}(T)$ obtained from the DMP equations and by averaging over $10^7$ Monte Carlo simulations. There is one randomly selected active node at the initial time and the dynamics is simulated for $t\!=\!5$ time steps.}
\label{fig:DMP-accuracy}
\end{figure}

To quantify the success of the spreading process, one may look for instance at the expected spread (the total number of infected nodes) at finial time horizon $T$, $\mathcal{S}(T)$, given by
\begin{equation}
\mathcal{S}(T)\!=\!\mathbb{E}\left[\sum_{i \in V}\mathds{1}[\sigma_{i}^{T}\!=\!I]\right] \!=\! \sum_{i \in V} P^{i}_{I}(T),
\end{equation}
where the expectation is taken with respect to the realization of the stochastic dynamics and $P^{i}_{I}(T)$ denotes the marginal probability of node $i$ to be found in the state $I$ at time $T$. The quantities $P^{i}_{S}(T)$ and $P^{i}_{R}(T)$ can be defined in a similar way for the susceptible and recovered states, respectively. Hence, it is important to understand how to compute approximately the marginal probabilities $P^{i}_{\sigma}(t)$ on a given network, $\sigma$ assuming the corresponding state; note that in the general case, an exact estimation of marginals in the SIR model is an NP-hard problem~\cite{shapiro2012finding}. We use the recently introduced Dynamic Message-Passing (DMP) equations~\cite{KarrerNewman2010,shrestha2014message,lokhov2015dynamic} which provide the estimates (asymptotically exact on sparse graphs) of the probabilities $P^{i}_{\sigma}(t)$ with a linear computational complexity in the number of edges and time steps. When applied to real-world loopy networks, the DMP algorithm typically yields a accurate prediction of the marginal probabilities as validated empirically~\cite{lokhov2015dynamic} for a large class of spreading models on real-world networks. In the~\emph{Methods} section, we provide an intuitive derivation of the corresponding DMP equations for the generalized SIR model. An example of the DMP performance on real-world networks is provided in the Figure~\ref{fig:DMP-accuracy}, where the method predictions are compared to values obtained through extensive Monte Carlo simulations of the SIR dynamics on a network of flights between major U.S. hubs (a detailed description of this data set is provided in the~\emph{Results} section and in the~\emph{Appendix~\ref{app:A}}). The accuracy of marginals estimation supports the use of the DMP equations at the core of our optimization algorithm.

\subsection{Optimization framework}

We formulate the dynamic allocation of resource as a general optimization problem with respect to an objective function $\mathcal{O}$ and a set of constraints associated with the budget of available resources $\mathcal{B}$, accessible values of control parameters $\mathcal{P}$, initial conditions $\mathcal{I}$ and the dynamical model equations $\mathcal{D}$. We employ the Lagrangian formulation of the constrained optimization problem:
\begin{equation}
\mathcal{L} \!=\! \underbrace{\mathcal{O}}_{\text{objective}}\hspace{-0.23cm} + \underbrace{\mathcal{B} +  \mathcal{P} + \mathcal{I} + \mathcal{D}}_{\text{constraints}}.
\label{eq:Lagrangian}
\end{equation}
Let us discuss the form of each term in the expression~\eqref{eq:Lagrangian}.

\underline{$\mathcal{O}$} - Many objective functions of interest relate to the delivered information~\emph{at particular times} defined for each node. So for the general case we define:
\begin{equation}
\mathcal{O} \!=\! \mathbb{E}\left[\sum_{i \in U}\mathds{1}[\sigma_{i}^{t_{i}}\!=\!I]\right] \!=\! \sum_{i} P^{i}_{I}(t_{i}),
\label{eq:targeting}
\end{equation}
where $t_{i}$ is the required activation time for node $i$ and the sum is over the subset of nodes $U \subset V$ that is required to be activated. We refer to this general formulation as the~\emph{targeting} problem. The popular problem of maximizing the total spread $\mathcal{S}(T)$ is a special case whereby $U \!=\! V$ and $t_{i} \!=\! T$ for all $i \in V$.

\underline{$\mathcal{B}$} - In many relevant situations, resources are not fully available at a given time, but rather become available on the fly, and their amount may vary across the time steps. For example, it takes some time to develop and produce the vaccines or the advertisement budget is allocated in stages depending on the success of the campaign. Hence, we define the budget constraints in the following form:
\begin{equation}
\sum_{i \in V} \nu_{i}(t) \!=\! B_{\nu}(t), \quad
\sum_{i \in V} \mu_{i}(t) \!=\! B_{\mu}(t),
\label{eq:Budget}
\end{equation}
where $B_{\nu}(t)$ and $B_{\mu}(t)$ denote the available total budget for the control parameters $\nu_{i}(t)$ (spontaneous infection) and $\mu_{i}(t)$ (recovery) at time $t$. The constraint $\mathcal{B}$ reads
\begin{equation}
\mathcal{B} \!=\! \sum_{t\!=\!0}^{T-1} \lambda^{\nu}_{B}(t) \left[\sum_{i \in V} \nu_{i}(t) - B_{\nu}(t) \right],
\label{eq:Budget_nu}
\end{equation}
with a similar expression for the parameters $\mu_{i}(t)$, where $\lambda^{\nu}_{B}(t)$ and $\lambda^{\mu}_{B}(t)$ are the associated Lagrange multipliers, respectively. Clearly, one is not forced to use the whole available budget at each time step; in this case, we assume that $B_{\nu}(t)$ and $B_{\mu}(t)$ are reallocated accordingly at subsequent time steps. However, in cases where~\emph{specific} targeting times are not required, using monotonicity arguments, it is easy to show that it is always advantageous to use all available budget fully at each step for maximizing the impact at a later stage. Allocation of budget at the initial time only corresponds to the optimal~\emph{seeding} problem.

\underline{$\mathcal{P}$} - In an unrestricted scenario, where all nodes are accessible, control parameters associated with node $i$, $\nu_{i}(t)$ and $\mu_{i}(t)$, may take arbitrary values from zero to one depending on total budget. However, in realistic situations access level to different nodes may differ: for example, only a subset $W \subseteq V$ of nodes may be controllable together with additional restrictions on parameter values. The parameter block $\mathcal{P}$ is introduced to enforce parameters $\nu_{i}(t)$ to take values in the range $[\underline{\nu_{i}^{t}},\overline{\nu_{i}^{t}}]$ at each time step. This can be accomplished with the help of ``barrier'' functions, widely used in constrained optimization, assuming the form
\begin{equation}
\mathcal{P} \!=\! \epsilon \sum_{t\!=\!0}^{T-1} \sum_{i \in V} \left( \log \left[ \nu_{i}(t)- \underline{\nu_{i}^{t}} \right] + \log \left[\overline{\nu_{i}^{t}} - \nu_{i}(t) \right] \right),
\label{eq:Parameters}
\end{equation}
where $\epsilon$ is a small regularization parameter chosen to minimize the impact on the objective $\mathcal{O}$ in the regime of allowed $\nu_{i}(t)$ values, away from the borders. An equivalent expression can be written for the constraints on the $\mu_{i}(t)$ values.

\underline{$\mathcal{I}$ and $\mathcal{D}$} - Finally, the constraints $\mathcal{I}$ and $\mathcal{D}$ enforce the given initial conditions and dynamics of the system via the associated Lagrange multipliers. For example, if no active individuals are present at initial time, then we set $P^{i}_{I}(0)\!=\! 0$ for all nodes using the constraint set $\mathcal{I}$; if some infected or recovered nodes are present, they assume an initial values 1 for the respective marginal probabilities. The set $\mathcal{D}$ encodes the evolution of the marginal probabilities with the DMP equations, as explained in the~\emph{Methods} section.

The extremization of the Lagrangian~\eqref{eq:Lagrangian} is done as follows. Variation of $\mathcal{L}$ with respect to the dual variables (Lagrange multipliers) results in the DMP equations starting from the given initial conditions, while derivation with respect to the primal variables (control and dynamic parameters) results in a second set of equations, coupling the Lagrange multipliers and the primal variable values at different times. We solve the coupled systems of equations by forward-backward propagation, a widely used method for learning and optimization in artificial neural networks~\cite{le1988theoretical}, detailed in the \emph{Methods} section. This method has a number of advantages compared to other localized optimization procedures such as gradient descent and its variants. In particular, it is simple to implement, is of modest computational complexity, does not require any adjustable parameters and is less prone to being trapped in local minima since the optimization is performed globally~\cite{saad1997globally}.

\begin{figure*}[hbt]
\begin{center}
\includegraphics[width=1.8\columnwidth]{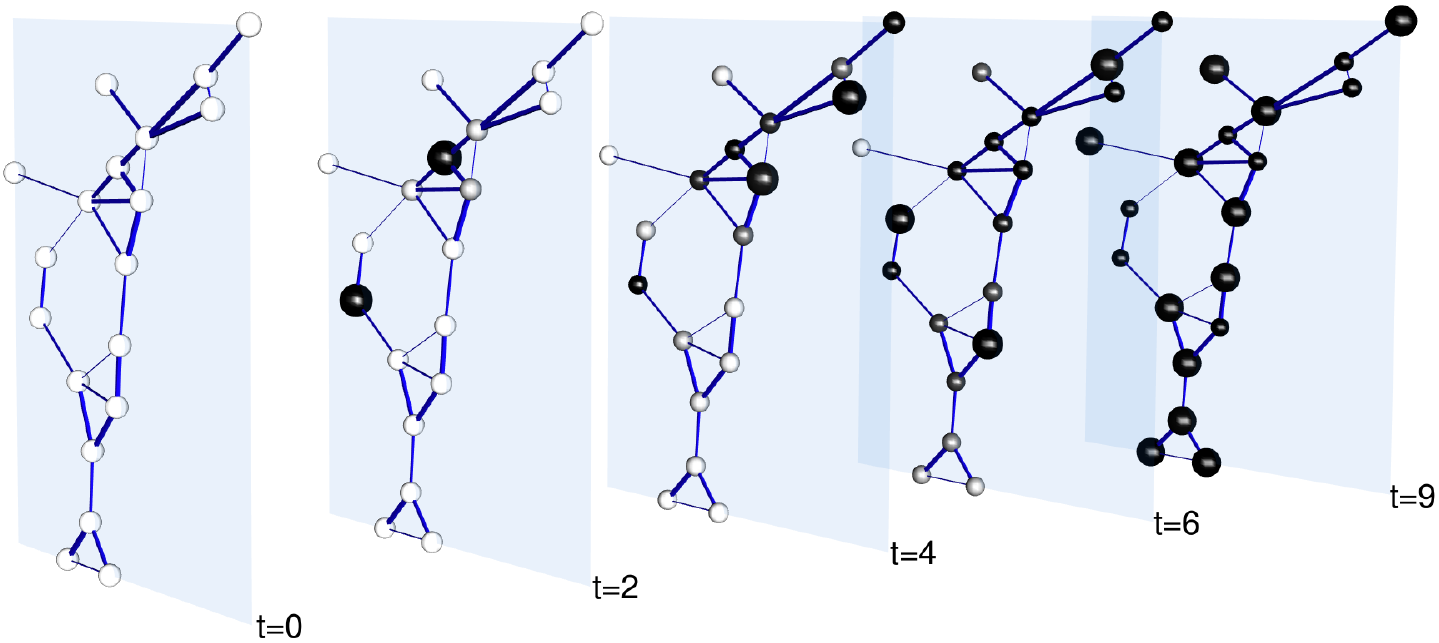}
\caption{{\bf Optimal targeting with the DMP algorithm on a small network of terrorists associations}. Edge thickness indicates the strength of the corresponding pairwise transmission probability $\alpha_{ij}$, generated uniformly at random in the interval $[0,1]$. The size of nodes relates to the time activation requirements: large nodes should be activated by the corresponding time. In this example, two chosen nodes should be activated at time $t\!=\!2$, another two nodes by time $t\!=\!4$, three particular nodes by time $t\!=\!6$ and all remaining nodes by time $t\!=\!9$; available budget for each time step has been fixed to $B_{\nu}(t)\!=\!0.1 \cdot N$. Color intensity (gradually from white to black) indicates the value of the marginal probabilities $P^{i}_{I}(t)$ which result from the dynamics using the optimal distribution of resources provided by the DMP algorithm. The visualization has been created using the MuxViz software~\cite{de2014muxviz}.}
\label{fig:Targeting}
\end{center}
\end{figure*}

\subsection{Targeting problem}

We first demonstrate the approach using the general targeting problem, one of the new features of the suggested framework. In this toy example, we consider disinformation spreading on a small network extracted from the study of terrorists associations~\cite{krebs2002mapping}. We assume that the spreading dynamics follows a particular case of the dynamical model with $\mu_{i}(t)\!=\!0~\forall~t~\mbox{and}~i \in V$, corresponding to the Susceptible-Infected, or SI model with controlled spontaneous transition to the informed state $I$ due to external influence via the control parameters $\nu_{i}(t)$. The activation of nodes is required in a predefined priority order, targeting selected nodes at specific times. The DMP-based optimization scheme converges to a unique optimal solution within a few forward-backward iterations as reported in Figure~\ref{fig:Targeting}. The resources are allocated dynamically over time such that the activation path meets the targeting requirements: $P^{i}_{I}(t_{i})>0.95$ is achieved at all nodes, with the majority of nodes targeted with probability one.

Targeting is quite a general task and can provide algorithms to solve a number of related problems. For instance, identifying the origin of the spreading process from measurements at sparsely located sensors at different times~\cite{Pinto2012} is a difficult problem that has been addressed by other approaches~\cite{lokhov2014inferring,Altarelli2014} but can be equally viewed as optimally allocating a budget at time zero in order to target the sensor nodes at specific times that correspond to the times when measurements were taken.

\begin{figure}[tb]
\begin{center}
\includegraphics[width=\columnwidth]{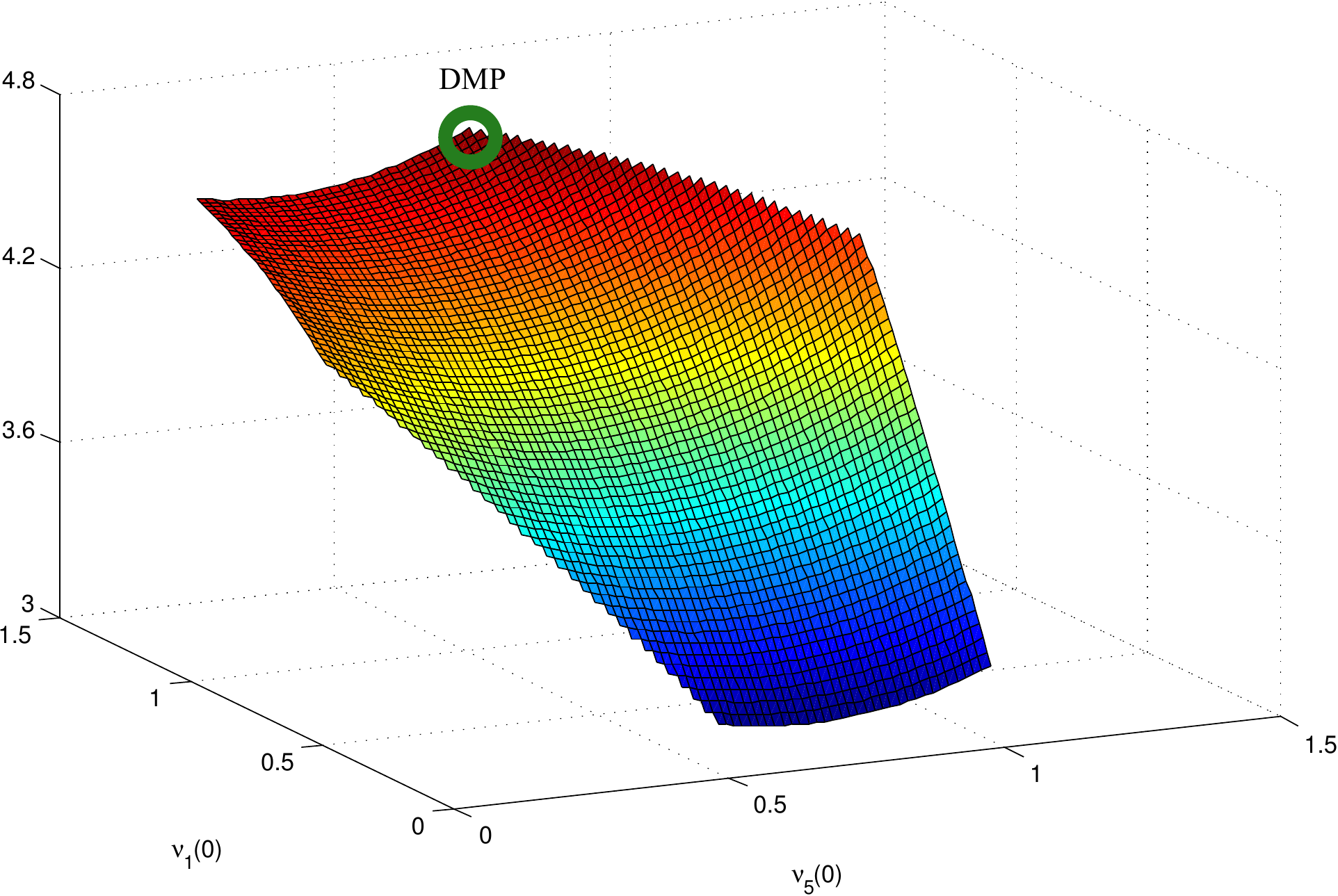}
\caption{{\bf Validation of the DMP algorithm on a small network of relations between Slovene parliamentary parties with an explicit evaluation of the objective function.} Edge weights have been renormalized such that the maximum pairwise mutual ``influence'' receives the value $\alpha^{\max}_{ij} \!=\! 0.5$ and other weights are distributed proportionally to the survey data. We assume that only three nodes in this network belong to a controllable set $W \!=\! \{1,3,5\}$, and the total campaigning budget $B_{\nu}(0) \!=\! 1.5$ so that $\nu_{3}(0) \!=\! 1.5 - \nu_{1}(0) - \nu_{5}(0)$. The plotted surface represents the the total informational ``influence'' $\mathcal{O}(\nu_{1}(0),\nu_{5}(0))$ as a function of two independent seeding control parameters $\nu_{1}(0)$ and $\nu_{5}(0)$, computed via a symbolic solution of the DMP equations. Color variations reflect the change of the value of the objective, changing gradually from the minimum (blue) to the maximum (red). The DMP algorithm correctly recovers the parameter values that maximize the total spread at the finial time (the corresponding solution is marked by a green circle).}
\label{fig:Symbolic}
\end{center}
\end{figure}

\subsection{Optimal seeding}

The majority of existing algorithms~\cite{pastor2002immunization,cohen2003efficient,holme2002attack,holme2004efficient,chen2008finding,kitsak2010identification,morone2015influence,altarelli2014containing,kempe2003maximizing} have been designed to solve the seeding problem -- finding an optimal set of nodes which would lead to the maximum number of activations at subsequent times. In the SI model of information spreading, even a single active node at initial time will ultimately lead to the activation of the whole connected network. However, a more interesting problem is the one of finding the best initial conditions which would lead to the maximum impact at finite time $T$. In the formulation involving control parameters $\nu_{i}(t)$, setting the initial conditions at time $t\!=\!1$ is equivalent to distributing the activation budget at time $t\!=\!0$ in the system where all nodes are at state $S$; optimal distribution of the budget at time $t\!=\!0$ would thus lead through spontaneous infection to the maximum spread $\mathcal{S}(T+1)$.

For demonstration and validation purposes, we first consider the small network of relations between Slovene parliamentary parties in 1994; links represent estimated similarity relations based on a sociological survey of the parliament members who were asked to estimate the distance between each pair of parties in the political space~\cite{doreian1996partitioning}. Given the total campaigning budget $B_{\nu}(0)$, the goal is to maximize the total informational ``influence'' at time $T$ by optimally distributing the ``lobbying'' budget at initial time. One advantage of this small test case is that we can check the validity of the scheme via an explicit symbolic solution of the DMP equations, obtaining a closed-form expression of the objective function $\mathcal{O} \!=\! \mathcal{S}(T+1)$ that represents the final information spread as a function of the independent control parameters; a more in-depth description of this procedure is given in the~\emph{Appendix~\ref{Sec:Spread_maximization}}. As in the previous example, the forward-backward optimization scheme in this case quickly converges to a unique optimal solution starting from an arbitrary initial values of the control parameters. The ground-truth optimal values of parameters can be established by a direct maximization of the objective function $\mathcal{O}$ plotted in the Figure~\ref{fig:Symbolic}. The optimal solution is in full agreement with the solution of the forward-backward iteration scheme up to the insignificant domain border perturbations due to the finite value of $\epsilon$, the regularization parameter that keeps values away from border values. This example validates our optimization procedure on this small scale problem.

Studies of the optimal seeding problem usually focus on a limited setting of homogeneous strength of links under deterministic dynamics and a search for the integer-valued deterministic budget deployment to specific nodes. To test the efficacy of the DMP-optimization approach on large scale instances we compare its performance to that of popular heuristics for this restricted setting. Although one should point out that our method addresses a broad range of problems and has not been optimized for this particular task, it is useful to assess its performance in the case where the structure of the ground-truth solution is known: for the deterministic SI spreading, it is clear that the initial distribution of seeds should target some combination of the high-degree nodes, and a number of well-performing centrality techniques~\cite{pastor2002immunization,kitsak2010identification,morone2015influence} are known to select the respective combination (see the~\emph{Appendix~\ref{Sec:Spread_maximization}} for a detailed discussion of methods used for comparison).

\begin{table*}[bt]
\caption{{\bf Comparison of the DMP algorithm for the seeding problem in the setting of deterministic dynamics with popular well-performing heuristics on various real-world networks.} The left of the Table provides topological information on the networks considered~\cite{vsubelj2011robust,bu2003topological,leskovec2007graph,boldi2004-ubicrawler}. On the right are presented values of the normalized total spread $\mathcal{S}(T)/N$ at time $T\!=\!3$ for the different algorithms: assignment to randomly-selected nodes, an adaptive version of the high-degree strategy of~\cite{pastor2002immunization} (HDA), k-shell decomposition~\cite{kitsak2010identification}, Collective Influence CI$_{l}$~\cite{morone2015influence} (with $l\!=\!2$ and $l\!=\!4$), uniform assignment and the DMP algorithm. For different test cases, solutions obtained by DMP span the range between delocalized and node-centric assignments and are on par with the best-performing centrality heuristics.}
\vspace{0.1cm}
\label{tab:real_networks}
 \setlength{\tabcolsep}{9.7pt}
 \renewcommand{\arraystretch}{1.1}
 \begin{tabular}{ l r r | c c c c c c c c c}
 Network  & {$N$} & {$M$}  & Random  & HDA  & k-shell  & CI{$_{2}$} & CI{$_{4}$} & Uniform & DMP \\
 \hline
 Road EU &1174 &1417 &0.305 &0.480 &0.160 &0.500 &0.468 &0.324 &0.513 \\
 Protein &2361 &6646 &0.736 &0.863 &0.769 &0.861 &0.838 &0.752 &0.856  \\
 US Power Grid &4941 &6594 &0.367 &0.602 &0.209 &0.605 &0.565 &0.397 &0.601 \\
 GR Collaborations &5242 &14484 &0.565 &0.644 &0.296 &0.660 &0.658 &0.634 &0.710 \\
 Internet &22963 &48436 &0.880 &0.998 &0.969 &0.996 &0.994 &0.891 &0.972 \\
 Web-sk &121422 &334419 &0.645 &0.833 &0.239 &0.751 &0.734 &0.699 &0.837 \\
 \hline
 \end{tabular}
\end{table*}

Table~\ref{tab:real_networks} presents the normalized total spread for some of the best-performing centrality measures and the DMP algorithm after $T\!=\!3$ time steps of the dynamics on different benchmark networks of various topologies and sizes. The transmission probabilities have been set to a uniform value $\alpha \!=\! 0.99$, and the total available seeding budget is equal to $B_{\nu}(0) \!=\! 0.05~N$. Note that the DMP-estimated marginals provide a natural and convenient measure for comparing the performance of different algorithms in the finite time horizon setting, especially on large graphs where running extensive Monte Carlo simulations is computationally prohibitive. Results presented in the Table~\ref{tab:real_networks} show that the DMP algorithm is close to the best-performing heuristics in all cases, showing a consistently good performance. Notice that our method does not rely explicitly on topological features such as targeting high-degree nodes, but instead explores a large space of parameters with impact on the full dynamic trajectory. This suggests that the DMP algorithm performs well also for more general dynamic resource allocation problems with heterogeneous couplings, for which other principled methods do not exist. In terms of computational complexity, solving the dynamics with DMP is linear in $T$ and $\vert E \vert$; the number of forward-backward iterations is typically small and can be controlled, as explained in the \emph{Methods} section. This compares well against the other algorithms even in the considered restricted setting where taking into account the dynamics is not required, and allows one to use the DMP approach for very large real-world networks. Additional implementation details and remarks are given in the~\emph{Appendix~\ref{Sec:Spread_maximization}}.

Given that the problem is NP-hard, it is not surprising that the optimization landscape is much more complex in the case of large networks due to a presence of multiple solutions with comparable costs; the forward-backward iteration scheme no longer converges to a unique optimum as in the case of small networks considered before. Instead, the algorithm ``jumps'' between local optima that representing different control-parameter distributions that obey the budget constraints~\eqref{eq:Budget}. This is an indication that it is arguably more appropriate to view the different seeding sets as a collective phenomenon, rather than assigning ``influence'' measure to individual nodes. In principle, several different initializations for $\nu_{i}(t)$ can be used to achieve the best solution; the results reported here correspond to the uniform starting values of the control parameters. Note that the initial distribution of $\nu_{i}(t)$ does not have to satisfy the budget condition~\eqref{eq:Budget}, but the solutions obey the constraint already after the first forward-backward iteration.

\subsection{Online mitigation of epidemic spreading}

To illustrate the suitability of the DMP algorithm to online deployment of resources in a dynamic setting with feedback we employ a prototypical example: developing an effective mitigation policy for confining an infectious disease -- a practical and challenging question of public concern. A SIR model with vaccination is an appropriate dynamic model in this case, where the $\nu_{i}(t)$ variables are set to zero, and the parameters $\mu_{i}(t)$ play the role of vaccination control, allowing the nodes to assume a protected state $R$. In contrast to the seeding problem, the initial conditions (origin of the epidemic) are specified in this setting and the vaccination budget has to be allocated dynamically according to the current state of the spreading process (monitored at each time step) in order to suppress the epidemic. The goal is to deploy the resources optimally so that the total number of infected nodes $\mathcal{S}(T)$ at the final time is minimized. The assumption of a time-distributed budget $B_{\mu}(t)$ is highly reasonable due to the restricted vaccine availability.

Previously developed real-time strategies for mitigating contagion on a given network~\cite{borgs2010distribute, drakopoulos2014efficient, scaman2015greedy} explored policies that are based on topological characteristics of the graph under the assumption of homogeneous transmission probabilities. The common denominator of existing approaches consists in local interventions which ensure the islanding of infected nodes. We generalize the methods~\cite{borgs2010distribute, drakopoulos2014efficient} to the case of heterogeneous transmission probabilities using a ``high-risk''~\cite{nian2010efficient} ranking of nodes according to their probability of getting infected at the next time step. This measure is defined in our case as
\begin{equation}
P^{t}_{i}(S \rightarrow I) \!=\! 1-\prod_{j \in \partial i}(1-\alpha_{ji}\mathds{1}[\sigma_{j}^{t}\!=\!I]),
\label{eq:high_risk}
\end{equation}
where $\partial i$ denotes the set of neighbors of node $i$. A reasonable local intervention strategy for benchmarking consists in distributing the vaccination budget to priority nodes with a high-risk measure~\eqref{eq:high_risk}. This algorithm will be referred to as the~\emph{greedy} strategy.

\begin{figure*}[bt]
\begin{center}
\includegraphics[width=2\columnwidth]{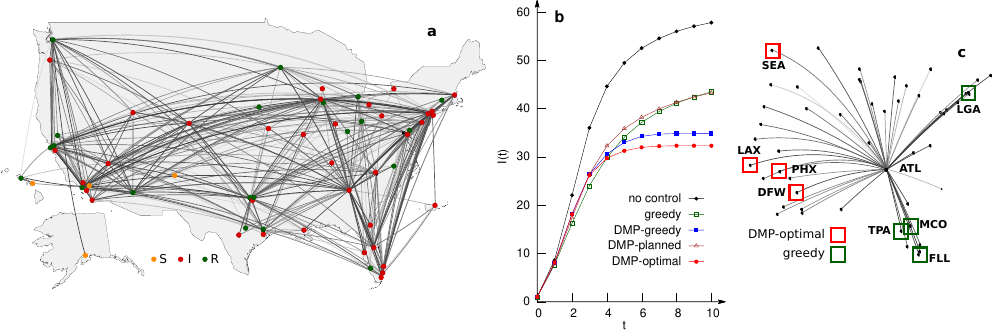}
\caption{{\bf Online mitigation of air-traffic mediated epidemic on the network of flights between major U.S. hubs.} (a) A geographical layout of the air transportation sub-network extracted from the BTS data~\cite{bts}. The transmission probabilities are indicated by the thickness of the corresponding edges, which is proportional to the aggregated traffic between airports. Different colors of airports (yellow, red and green) represent an outcome of a single realization of the spreading dynamics (nodes in the susceptible, infected and recovered states, respectively) under the DMP-optimal policy. (b) Comparisons of mitigation strategies showing the average number of infected sites as a function of time, averaged over $100$ random realizations of the dynamics, as a result of applying different policies. In the simulations the epidemic starts at the largest airport hub of Atlanta; a budget of $B_{\mu}(t) \!=\! 0.5~N$ is available at each time step and the objective is to suppress the epidemic by $T\!=\!10$. The DMP-optimal algorithm demonstrates the best performance in the number of infected nodes at time $T$. (c) An illustration of a radically different decisions taken by the DMP-optimal and greedy algorithms already at the first step of the optimization: the greedy policy chooses to vaccinate nodes which are most ``in danger'' at the next time step, while the decision done by the DMP-optimal scheme takes into account the forecasted evolution of the dynamics.}
\label{fig:airflights}
\end{center}
\end{figure*}

Several policies can be conceived using the DMP optimization framework. As a reference, we consider the~\emph{planned} deployment of resources which does not take into account feedback from an actual realization of the process, but merely follows the solution of the dynamic resource allocation problem with a specified initial condition. Two other closed-loop strategies take into account the real-time information on the spreading process, using the seeding formulation as a subroutine: (a) The first, termed \emph{DMP-greedy}, is close in spirit (but differs in the algorithmic implementation, based here on the DMP optimization framework) to the greedy algorithm and uses the current state of the epidemic as the initial condition, aiming to minimize the spread at the next time step only. (b) The second utilizes the full power of the DMP framework by exploiting the up-to-date information available to reinitialize the dynamics at each time step $t$ to allocate the resources at the next time step $t+1$, by running the optimization procedure for the remaining $T-t$ time steps. This~\emph{DMP-optimal} policy is similar to the planned strategy, but takes advantage of the new information available from the realization of the process.

We compare these strategies for the case of infection spreading mediated by air traffic, which has been recognized to play an important role in recent world's pandemics~\cite{Tatem2006}. As a particular example, we study the real-world transportation network of busiest flight routes between major U.S.~airports, extracted from the Bureau of Transportation Statistics data (BTS)~\cite{bts} and depicted in Figure~\ref{fig:airflights}~(a). We employ a plausible assumption that the infection transmission probability associated with a link between airports is proportional to the number of passengers carried along this route, see the~\emph{Appendix~\ref{app:C}} for a detailed description of the network and data used. The ``vaccination'' interventions on this network can be interpreted as quarantine measures taken in different airports using the updates on the new infected cases. In the simulations, we assume that the epidemic starts at the largest airport hub of Atlanta.

The comparison of different mitigation algorithms is given in Figure~\ref{fig:airflights}~(b), showing the average number of infected sites as a function of time under different mitigation strategies. As expected, the DMP-optimal scheme represents the best performing policy, which leads to stabilization of the expected number of infected nodes by $t\!=\!6$, at a lower level compared to the greedy algorithm that optimizes the spread at the next time step only. Notice that on a short time scale, the greedy algorithm has a slightly better performance, which represents a typical situation when localized and immediate optimal decisions lead ultimately to sub-optimal global optimization results.

\section{Discussion}

We introduced an efficient, versatile and principled optimization framework for solving dynamic resource allocation problems in spreading processes, which allows for the synthesis of previously studied settings using a general targeting formulation. This probabilistic framework allows for the study of problems that involve a finite-time horizon, which requires an explicit solution of the dynamics, the targeting of specific nodes at given times and it accommodates scenarios where only a subset of the nodes is accessible. This is done in our scheme using the DMP equations for spreading processes. Although in this work we focused on the examples involving the discrete-time generalized SIR model, the approach can be straightforwardly applied to the case of continuous dynamics (the continuous formulation is expounded in the~\emph{Appendix~\ref{sec:Continuous}}) and to other spreading models, such as cascading and threshold models as well as rumor dynamics~\cite{lokhov2015dynamic}. Another possible application area of the present framework relates to systems defined on temporal graphs, where network dynamics can be encoded into the time-dependent coefficients $\alpha_{ij}(t)$ within the existing framework.

Although we showed that the method can be employed in the case where transmission probabilities are uniform and only the detailed topology of the network is known, its major advantage consists in the ability to incorporate efficiently detailed information on transmission probabilities when such prior information is available, or can be either estimated (as in the examples of the Slovene political parties or flight transportation networks) or learned from observations of the dynamics~\cite{lokhov2015efficient}.

The optimization method used is interesting in itself being based on changes to the entire trajectory instead of taking incremental improvement steps in the direction of the gradient; thus, the suggested algorithm results in large steps and arguably explores more effectively the parameter space. In spite of the global budget constraints involving all network nodes the resulting message-passing scheme is fast and distributed, requiring a number of operations which grows linearly in time and with respect to the number of edges in the network. An attractive property of the suggested framework is its versatility: instead of optimizing the spread given a fixed budget one can minimize the budget needed to meet certain requirements on the spread, imposed as a constraint in the Lagrangian formulation. Another interesting scenario is the optimization over the spreading parameters $\alpha_{ij}$: this formulation is useful in the design of technological networks or for mitigation of an epidemic by removing and adding links in the graph. Finally, it would be interesting to apply the presented optimization scheme to the percolation-type equations describing the asymptotic $T \to \infty$ limit of the spreading dynamics with heterogeneous couplings. Work on these research directions are underway.

\section{Methods}

\subsection{Dynamic message-passing equations} Dynamic message-passing belongs to the family of algorithms derived using the cavity method of statistical physics and may be given an interpretation of passing messages along the graph edges. The obtained marginals are exact on tree graphs, and asymptotically exact on sparse random networks. We provide an intuitive derivation of the DMP equations for the adopted generalized SIR model, defined by \eqref{eq:SIR_rules_StoI_interaction} and \eqref{eq:SIR_rules_StoI_spontaneous}. On a given instance of a network, these equations allow one to compute the marginal probability distributions $P^{i}_{\sigma}(t)$, where $\sigma \in \{S,I,R\}$ denotes the node state. The first key equation reads:
\begin{equation}
P_{S}^{i}(t)\!=\!P_{S}^{i}(0)\left(\prod_{t'\!=\!0}^{t-1}(1-\nu_{i}(t'))(1-\mu_{i}(t'))\right)\prod_{k\in \partial i}\theta^{k \rightarrow i}(t). \label{eq:SImarginals:S}
\end{equation}
It states the probability of node $i$ to be susceptible at time $t$ and is equal to the probability that $i$ was in the $S$ state at initial time $P_{S}^{i}(0)$ and remained so until time $t$. It neither changed states by following the $\nu$ and $\mu$ mechanisms (in brackets), nor by being infected by a neighbor (final term on right); the dynamic message $\theta^{k \rightarrow i}(t)$ has a meaning of the probability
that node $k$ did not pass an activation message to node $i$ until time $t$. Strictly speaking, Eq.~\eqref{eq:SImarginals:S} is only valid on a tree graph; only in this case $\theta^{k \rightarrow i}(t)$ are independent for all $k \in \partial i$, so that the corresponding probability is factorized as in~\eqref{eq:SImarginals:S}. However, in practice the decorrelation assumption holds to a good precision even on general networks, even with small loops, see~\cite{lokhov2015dynamic} for in-depth discussions and supporting numerical experiments. The quantities $\theta^{k \rightarrow i}(t)$ are updated as follows:
\begin{equation}
\theta^{k \rightarrow i}(t)\!=\!\theta^{k \rightarrow i}(t\!-\!1)
-\alpha_{ki}\phi^{k \rightarrow i}(t\!-\!1),
\end{equation}
which corresponds to the fact that $\theta^{k \rightarrow i}(t)$ can only decrease if an activation signal is passed along the directed link $(ki)$; the corresponding probability equals the product of $\alpha_{ki}$ and the dynamic variable $\phi^{k \rightarrow i}(t\!-\!1)$, which has a meaning of the probability that node $k$ is in the state $I$ at time $t$, but has not infected node $i$ until time $t\!-\!1$. To simplify further explanations we introduce the dynamic messages $P_{S}^{k \rightarrow i}(t)$, $P_{I}^{k \rightarrow i}(t)$ and $P_{R}^{k \rightarrow i}(t)$, which denote the probabilities that node $k$ is found at time $t$ in the states $S$, $I$ or $R$, respectively, conditioned on node $i$ remaining in state $S$. Alternatively, these variables can be thought of as the probabilities of $k$ being susceptible, infected or recovered on a \emph{cavity graph}, on which node $i$ has been removed. Formally,
\begin{equation}
P_{S}^{k \rightarrow i}(t)\!=\!P_{S}^{k}(0)\Bigg(\prod_{t'\!=\!0}^{t\!-\!1}(1\!-\!\nu_{k}(t'))(1\!-\!\mu_{k}(t'))\Bigg)\prod_{l\in \partial k \backslash i}\theta^{l \rightarrow k}(t),
\end{equation}
which coincides with the expression~\eqref{eq:SImarginals:S}, except that $\theta^{i \rightarrow k}(t)$ is not included in the product on the right ($\partial k \backslash i$ denotes the set of neighbors of $k$ without $i$). We also have
\begin{equation}
P_{R}^{k \rightarrow i}(t) \!=\! P_{R}^{k \rightarrow i}(t\!-\!1) + \mu_{k}(t\!-\!1)P_{S}^{k \rightarrow i}(t),
\end{equation}
which expresses the monotonic increase of $P_{R}^{k \rightarrow i}(t)$ at each time step with the probability $\mu_{k}(t\!-\!1)P_{S}^{k \rightarrow i}(t)$, and
\begin{equation}
P_{I}^{k \rightarrow i}(t) \!=\! 1 \!-\! P_{S}^{k \rightarrow i}(t) \!-\! P_{R}^{k \rightarrow i}(t)
\end{equation}
due to the properties of linked probabilities. We are now ready to formulate the last relation which leads to the closure of the system of message-passing equations. The evolution of the message $\phi^{k \rightarrow i}(t)$ reads:
\begin{equation}
\phi^{k \rightarrow i}(t)\!=\!(1\!-\!\alpha_{ki})\phi^{k \rightarrow i}(t\!-\!1) \!+\! \Delta P_{I}^{k \rightarrow i}(t\!-\!1)
\label{eq:phi}
\end{equation}
where $\Delta P_{I}^{k \rightarrow i}(t\!-\!1) \equiv P_{I}^{k \rightarrow i}(t) \!- \!P_{I}^{k \rightarrow i}(t\!-\!1)$. The physical meaning of equation~\eqref{eq:phi} is as follows: $\phi^{k \rightarrow i}(t)$ decreases if the activation signal is actually transmitted (first term) and increases if node $k$ transitions to the state $I$ at the current time step. Equations \eqref{eq:SImarginals:S}-\eqref{eq:phi} can be iterated in time starting from the given initial conditions $\{P_{S}^{i}(0),P_{I}^{i}(0),P_{R}^{i}(0)\}_{i \in V}$, with
\begin{equation}
\theta^{i \rightarrow j}(0)\!=\!1, \quad
\phi^{i \rightarrow j}(0)\!=\!\delta_{\sigma_{i}^{0},I}\!=\!P_{I}^{i}(0).
\label{eq:SIequations:initial_conditions_phi}
\end{equation}

The marginals $P_{S}^{i}(t)$ used throughout the text are obtained using \eqref{eq:SImarginals:S}, while  $P_{I}^{i}(t)$ and $P_{R}^{i}(t)$ are computed via
\begin{align}
P_{R}^{i}(t)&\!=\!P_{R}^{i}(t\!-\!1) \!+\! \mu_{i}(t\!-\!1)P_{S}^{i}(t\!-\!1),\label{eq:SImarginals:I}
\\
P_{I}^{i}(t)&\!=\!1\!-\!P_{S}^{i}(t)\!-\!P_{R}^{i}(t). \label{eq:SImarginals:R}
\end{align}
The computational complexity of the DMP equations for solving the dynamics up to time $T$ is given by $O(\vert E \vert T)$, where $\vert E \vert$ is the number of edges in the graph, which makes them scalable to sparse networks with millions of nodes.

\subsection{Enforcing dynamical constraints and backward equations} The dynamics $\mathcal{D}$ and initial conditions $\mathcal{I}$ constraints are enforced in a similar way to that of $\mathcal{P}$ and the budget $\mathcal{B}$ constraints in Eqs.~\eqref{eq:Budget_nu} and \eqref{eq:Parameters}. To each generic dynamic variable $\mathcal{\xi}^{i}(t)$ and message $\mathcal{\chi}^{k \rightarrow i}(t)$ we associate the corresponding Lagrange multipliers $\lambda^{\xi}_{i}(t)$ and $\lambda^{\chi}_{k \rightarrow i}(t)$ which enforce the relation between dynamic variables at subsequent times. For instance, the evolution of the quantities $\{P_{R}^{i}(t)\}_{i \in V}$ in the Lagrangian $\mathcal{L}$ is enforced via the term
\begin{equation*}
\sum_{i \in V}\sum_{t\!=\!0}^{T\!-\!1}\lambda^{R}_{i}(t\!+\!1)\left[ P_{R}^{i}(t\!+\!1) \!-\! P_{R}^{i}(t) \!-\! \mu_{i}(t)P_{S}^{i}(t) \right].
\end{equation*}

Variation with respect to the dual variables $\lambda^{\xi}_{i}(t)$ and $\lambda^{\chi}_{k \rightarrow i}(t)$ returns the \emph{forward} DMP equations~\eqref{eq:SImarginals:S}-\eqref{eq:SImarginals:R}, while setting to zero the derivative of $\mathcal{L}$ with respect to the primal dynamic variables yields the relations between the Lagrange multipliers at subsequent times, which we interpret as the \emph{backward} dynamic equations in our scheme. Similarly to~\eqref{eq:SImarginals:S}-\eqref{eq:phi}, the backward equations have a distributed message-passing structure with linear computational complexity $O(\vert E \vert T)$, and are used to update the values of control parameters $\nu_{i}(t)$ and $\mu_{i}(t)$ at each iteration, taking into account the budget requirements~\eqref{eq:Budget}. Specifically, initializing the control parameters $\nu_{i}(t)$ and $\mu_{i}(t)$ to some arbitrary values (e.g., uniform over all nodes and times), we first propagate the DMP equations~\emph{forward} in time, up to the horizon $T$; then, using the existing primal parameter values we fix end-point conditions for the dual parameters and propagate the equations for the dual parameters~\emph{backward} in time, updating the control parameters respecting the budget and variation constraints. These two steps are iterated for a predefined number of times or until global convergence of the process.

In the large-scale problems, where the algorithm explores the space of parameters by hopping from one solution to another, we choose a simple strategy: we run the forward-backward algorithm for several iterations for a range of values of the regularization parameter $\epsilon$ which appears in the $\mathcal{P}$ block, and keep track of the best local optimum which provides the solution to the optimization problem after a maximum number of iterations (kept below the desired threshold which determines the computational complexity) is reached. The choice of $\epsilon$ impacts on the type of solution obtained: larger values of $\epsilon$ correspond to solutions where the budget is disseminated more uniformly across nodes, while smaller values lead to weight concentration on particular nodes. Depending on the application and the level of control over nodes, one type of solution can be preferred to another; this flexibility represents an attractive feature of the DMP algorithm. An explicit form of the Lagrangian for the problems considered in this work together with additional details is given in the~\emph{Appendices \ref{Sec:Spread_maximization} and \ref{app:C}}.

\begin{acknowledgments}
\vspace{-0.5cm}
The authors are grateful to M.~Chertkov, S.~Misra and M.~Vuffray for fruitful discussions and valuable comments. A.Y.~Lokhov acknowledges support from the LDRD Program at Los Alamos National Laboratory by the National Nuclear Security Administration of the U.S.~Department of Energy under Contract No.~DE-AC52-06NA25396. D.~Saad acknowledges support from the Leverhulme Trust RPG-2013-48.
\end{acknowledgments}

\onecolumngrid
\appendix
\small


\section{Model and DMP equations}
\label{app:A}

\begin{figure}[!h]
\begin{center}
\includegraphics[width=0.25\columnwidth]{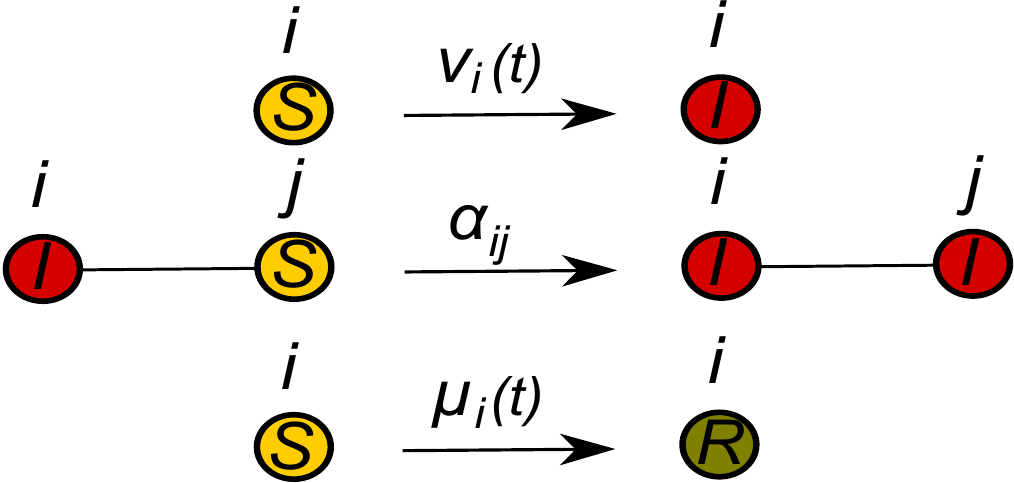}
\caption{Transition diagram summarizing dynamic rules in the generalized SIR model considered in this work. In the case of a simultaneous transition of a susceptible node (marked in yellow) to the states $I$ (infected - red) and $R$ (recovered - green) at time $t$, it assumes a recovered state at the next time step.}
\label{fig:S1}
\end{center}
\end{figure}

All models considered in the main text represent variants of a SIR model with possible spontaneous infection and vaccination transitions. We refer to this model as \emph{generalized} SIR model, because usually the spontaneous $S \rightarrow I$ and $S \rightarrow R$ transitions are not considered; instead, the standard SIR model contains a spontaneous recovery transition $I \rightarrow R$ which was not relevant for the examples considered in the paper and therefore was ignored. However, the inclusion of this transition in the DMP equations is very easy and has been done in \cite{lokhov2015dynamic}. Hence, the model we consider here is defined as follows: at each time step $t$, the transitions from the state $S$ to $I$ and $R$ occur with the probabilities summarized in the Figure~\ref{fig:S1} for individual nodes and edges. In the discrete time setting considered throughout the work, it may occur that both transitions to $I$ and $R$ states are realized at the same time; in this case, we assume that the transition to the $R$ state effectively takes place. Note that in the continuous time setting (see Section~\ref{sec:Continuous}), this effect is of a second order in the discretization step $dt$, and hence this tie-breaking rule is not required for sufficiently small $dt$.

The DMP equations associated with this model, as well as the intuition behind them are described in the {\it Methods} section of the main text. We repeat them here for consistency:

\begin{align}
P_{S}^{k \rightarrow i}(t)&=P_{S}^{k}(0)\Bigg(\prod_{t'=0}^{t-1}(1-\nu_{k}(t'))\Bigg)\Bigg(\prod_{t'=0}^{t-1}(1-\mu_{k}(t'))\Bigg)\prod_{l\in \partial k \backslash i}\theta^{l \rightarrow k}(t), \label{app:eq:SIequations:S}
\\
P_{R}^{k \rightarrow i}(t)&=P_{R}^{k \rightarrow i}(t-1)+\mu_{k}(t-1)P_{S}^{k \rightarrow i}(t-1),
\\
\theta^{k \rightarrow i}(t)&=\theta^{k \rightarrow i}(t-1)
-\alpha_{ki}\phi^{k \rightarrow i}(t-1),
\\
\phi^{k \rightarrow i}(t)&=(1-\alpha_{ki})\phi^{k \rightarrow i}(t-1) + \left[ \left( P_{S}^{k \rightarrow i}(t-1)-P_{S}^{k \rightarrow i}(t) \right) - \left( P_{R}^{k \rightarrow i}(t)-P_{R}^{k \rightarrow i}(t-1) \right) \right].
\label{app:eq:SIequations:phi}
\end{align}

The initial conditions are
\begin{align}
\theta^{i \rightarrow j}(0)=1, \quad \quad 
\phi^{i \rightarrow j}(0)=\delta_{\sigma_{i}^{0},I}=P_{I}^{i}(0)=1-P_{S}^{i}(0).
\label{app:eq:SIequations:initial_conditions}
\end{align}

The marginal probabilities for nodes to be in the states $S$ or $I$ at time $t$ are computed via
\begin{align}
P_{S}^{i}(t)&=P_{S}^{i}(0)\left(\prod_{t'=0}^{t-1}(1-\nu_{i}(t'))\right)\left(\prod_{t'=0}^{t-1}(1-\mu_{i}(t'))\right)\prod_{k\in \partial i}\theta^{k \rightarrow i}(t), \label{app:eq:SImarginals:S}
\\
P_{R}^{i}(t)&= P_{R}^{i}(t-1) + \mu_{i}(t-1)P_{S}^{i}(t-1),\label{app:eq:SImarginals:R}
\\
P_{I}^{i}(t)&=1-P_{S}^{i}(t)-P_{R}^{i}(t).\label{app:eq:SImarginals:I}
\end{align}

As explained in the main text, the optimization problem is stated in the form of a Lagrangian to be extremized:
\begin{equation}
\mathcal{L} = \underbrace{\mathcal{O}}_{\text{objective}} + \underbrace{\mathcal{B} + \mathcal{D} + \mathcal{I} + \mathcal{P}}_{\text{constraints}},
\label{app:eq:Lagrangian}
\end{equation}
where $\mathcal{O}$ is the objective function one would like to maximize, and $\mathcal{B}$, $\mathcal{D}$, $\mathcal{I}$ and $\mathcal{P}$ correspond to the constraints representing budget, dynamics, initial conditions and limits on the parameters, respectively. Below we discuss in detail how to obtain an approximate solution to the optimal control problem using this framework, for the examples of the spread maximization or minimization under different constraints. In all examples, we assume that budget constraints for spontaneous infections or vaccination take a global form over a subset $W \subseteq V$ of nodes in the network, and are specified at each time step:
\begin{align}
\sum_{i \in W} \nu_{i}(t) = B_{\nu}(t), \quad \quad
\sum_{i \in W} \mu_{i}(t) = B_{\mu}(t).
\label{app:eq:Budget_nu}
\end{align}
Without loss of generality and for the sake of simplicity, in the equations below we assume that all nodes in the network are controllable, $W = V$; this case is the hardest in terms of the optimization procedure. Extending the derivation to the general case of any subset $W$ is straightforward (an  illustration for the case $W \neq V$ will be given in the Section~\ref{sec:Seeding_tests} below).

\section{Maximizing information spread under special targeting policy}
\label{Sec:Spread_maximization}

In this section, we write the detailed form of the Lagrangian for the targeting problem, explained in the main text: we assume that for each node $i \in V$ the activation is required at a predefined time $t_{i}$. We derive the corresponding forward and backward equations for the particular case of the generalized SIR model without a vaccination transition, i.e. assuming $\mu_{i}(t)=0$ for all $i \in V$ and $t \in [0,T-1]$.

\subsection{Lagrangian formulation}
The Lagrangian in this case takes the following form:

\begin{align}
\notag
\mathcal{L}&=\underbrace{\sum_{i\in V}\left(1-P^{i}_{S}(t_{i})\right)}_{\mathcal{O}}
+\underbrace{\sum_{t=0}^{T-1} \lambda^{\nu}_{B}(t) \left[\sum_{i \in V} \nu_{i}(t) - B_{\nu}(t) \right]}_{\mathcal{B}}
+\underbrace{\epsilon \sum_{t=0}^{T-1} \sum_{i \in V} \left( \log \left[ \nu_{i}(t)- \underline{\nu_{i}^{t}} \right] + \log \left[\overline{\nu_{i}^{t}} - \nu_{i}(t) \right] \right)}_{\mathcal{P}}
\\
\notag
&\left.
\begin{aligned}
&+\sum_{i\in V}\sum_{t=0}^{T-1}\lambda^{S}_{i}(t+1)\left[P_{S}^{i}(t+1)-P_{S}^{i}(t)(1-\nu_{i}(t))\prod_{k \in \partial i}\frac{\theta^{k \rightarrow i}(t+1)}{\theta^{k \rightarrow i}(t)}\right]
\\
&+\sum_{(ki)\in E}\sum_{t=0}^{T-1}\lambda^{S}_{k \rightarrow i}(t+1)\left[P_{S}^{k \rightarrow i}(t+1)-P_{S}^{k \rightarrow i}(t)(1-\nu_{k}(t))\prod_{l\in \partial k \backslash i}\frac{\theta^{l \rightarrow k}(t+1)}{\theta^{l \rightarrow k}(t)}\right]
\\
&+\sum_{(ki)\in E}\sum_{t=0}^{T-1}\lambda^{\theta}_{k \rightarrow i}(t+1)\left[\theta^{k \rightarrow i}(t+1)-\theta^{k \rightarrow i}(t)
+\alpha_{ki}\phi^{k \rightarrow i}(t)\right]
\\
&+\sum_{(ki)\in E}\sum_{t=0}^{T-1}\lambda^{\phi}_{k \rightarrow i}(t+1)\Big[\phi^{k \rightarrow i}(t+1)-(1-\alpha_{ki})\phi^{k \rightarrow i}(t)
-P_{S}^{k \rightarrow i}(t)+P_{S}^{k \rightarrow i}(t+1)\Big]
\end{aligned}
\right\rbrace\mathcal{D}
\\
\notag
&\left.
\begin{aligned}
&+\sum_{i\in V}\lambda^{S}_{i}(0)\left[P^{i}_{S}(0)-1+\delta_{\sigma_{i}^{0},I}\right]+\sum_{(ki)\in E}\lambda^{S}_{k \rightarrow i}(0)\left[P_{S}^{k \rightarrow i}(0)-1+\delta_{\sigma_{k}^{0},I}\right]
\\
&+\sum_{(ki)\in E}\lambda^{\theta}_{k \rightarrow i}(0)\left[\theta^{k \rightarrow i}(0)-1\right]+\sum_{(ki)\in E}\lambda^{\phi}_{k \rightarrow i}(0)\left[\phi^{k \rightarrow i}(0)-\delta_{\sigma_{k}^{0},I}\right].
\end{aligned}
\right\rbrace\mathcal{I}
\end{align}
In this expression, the dual variables $\lambda^{S}_{i}(t)$, $\lambda^{S}_{k \rightarrow i}(t)$, $\lambda^{\theta}_{k \rightarrow i}(t)$ and $\lambda^{\phi}_{k \rightarrow i}(t)$ in ${\cal D}$ and ${\cal I}$ enforce the dynamics given by the DMP equations as well as the initial conditions at time zero, while $\lambda^{\nu}_{B}(t)$ is the corresponding Lagrange multiplier for the budget constraint in ${\cal B}$. The parameter constraint $\mathcal{P}$ has a form of the logarithmic barrier function, often used in constrained optimization, forcing each of the parameters $\nu_{i}(t)$ to take vales inside the interval $[\underline{\nu_{i}^{t}},\, \overline{\nu_{i}^{t}}]$; for sufficiently small value of a positive coefficient $\epsilon$, this regularization has a negligible impact on the objective function away from the extreme values.

Note that in the formulation above we have made an implicit assumption that $\alpha_{ki} < 1$ for all $(ki) \in E$, so that the ratios $\frac{\theta^{k \rightarrow i}(t+1)}{\theta^{k \rightarrow i}(t)}$ are correctly defined. In most situations, the case $\alpha_{ki} = 1$ for some $(ki) \in E$ is somewhat trivial, because there is no need for control for the nodes adjacent to these links. This assumption can be easily avoided by introducing other auxiliary variables in order to decouple $\theta^{k \rightarrow i}(t)$ at different times, which results in a slightly more complicated formulation.

\subsection{Forward and backward equations}

Once the form of the Lagrangian is established, we use a standard derivation with respect to primal and dual variables. The variation with respect to the multipliers $\lambda^{S}_{i}(t)$, $\lambda^{S}_{k \rightarrow i}(t)$, $\lambda^{\theta}_{k \rightarrow i}(t)$ and $\lambda^{\phi}_{k \rightarrow i}(t)$ give us back the direct DMP equations~\eqref{app:eq:SIequations:S}-\eqref{eq:SImarginals:I}, while the variation with respect to $\lambda^{\nu}_{B}(t)$ yields the cost constraint~\eqref{app:eq:Budget_nu}. Setting the derivatives with respect to the primal variables to zero leads to the following set of dual equations:

\begin{align}
&\partial \mathcal{L} / \partial \phi^{k \rightarrow i}(t) = \left[\alpha_{ki}\lambda^{\theta}_{k \rightarrow i}(t+1)-(1-\alpha_{ki})\lambda^{\phi}_{k \rightarrow i}(t+1)\right]\mathds{1}[t\neq T]+\lambda^{\phi}_{k \rightarrow i}(t)=0,
\label{eq:variation_phi}
\\
\notag
&\partial \mathcal{L} / \partial P_{S}^{k \rightarrow i}(t) = \lambda^{S}_{k \rightarrow i}(t)+\lambda^{\phi}_{k \rightarrow i}(t)
\\
&-\left[\lambda^{S}_{k \rightarrow i}(t+1)(1-\nu_{k}(t))\prod_{l \in \partial k \backslash i}\frac{\theta^{l \rightarrow k}(t+1)}{\theta^{l \rightarrow k}(t)}+\lambda^{\phi}_{k \rightarrow i}(t+1)\right]\mathds{1}[t\neq T]=0,
\label{eq:variation_P_S}
\\
\notag
&\partial \mathcal{L} / \partial \theta^{k \rightarrow i}(t) = \lambda^{\theta}_{k \rightarrow i}(t)-\lambda^{\theta}_{k \rightarrow i}(t+1)\mathds{1}[t\neq T]
\\
\notag
&-\sum_{l \in \partial i \backslash k}\lambda^{S}_{i \rightarrow l}(t)P^{i \rightarrow l}_{S}(t-1)(1-\nu_{i}(t-1))\frac{1}{\theta^{k \rightarrow i}(t-1)}\prod_{m\in \partial i \backslash \{k,l\}}\frac{\theta^{m \rightarrow i}(t)}{\theta^{m \rightarrow i}(t-1)}\mathds{1}[t \neq 0]
\\
\notag
&+\sum_{l \in \partial i \backslash k}\lambda^{S}_{i \rightarrow l}(t+1)P^{i \rightarrow l}_{S}(t)(1-\nu_{i}(t))\frac{1}{\theta^{k \rightarrow i}(t)}\prod_{m\in \partial i \backslash l}\frac{\theta^{m \rightarrow i}(t+1)}{\theta^{m \rightarrow i}(t)}\mathds{1}[t \neq T]
\\
\notag
&-\lambda^{S}_{i}(t)P^{i}_{S}(t-1)(1-\nu_{i}(t-1))\frac{1}{\theta^{k \rightarrow i}(t-1)}\prod_{m\in \partial i \backslash k}\frac{\theta^{m \rightarrow i}(t)}{\theta^{m \rightarrow i}(t-1)}\mathds{1}[t \neq 0]
\\
&+\lambda^{S}_{i}(t+1)P^{i}_{S}(t)(1-\nu_{i}(t))\frac{1}{\theta^{k \rightarrow i}(t)}\prod_{m\in \partial i }\frac{\theta^{m \rightarrow i}(t+1)}{\theta^{m \rightarrow i}(t)}\mathds{1}[t \neq T]=0.
\label{eq:variation_theta}
\\
&\partial \mathcal{L} / \partial P_{S}^{i}(t) = -\mathds{1}[t = t_{i}]+\lambda^{S}_{i}(t)-\lambda^{S}_{i}(t+1)(1-\nu_{i}(t))\prod_{k \in \partial i}\frac{\theta^{k \rightarrow i}(t+1)}{\theta^{k \rightarrow i}(t)}\mathds{1}[t \neq T]=0.
\label{eq:variation_P_S_marginal}
\end{align}
Finally, the variation with respect to the parameters $\nu_{i}(t)$ for $t<T$ gives
\begin{align}
\notag
\partial \mathcal{L} / \partial \nu_{i}(t) = \lambda^{\nu}_{B}(t) + \sum_{l \in \partial i }\lambda^{S}_{i \rightarrow l}(t+1)P^{i \rightarrow l}_{S}(t)\prod_{m\in \partial i \backslash l}&\frac{\theta^{m \rightarrow i}(t+1)}{\theta^{m \rightarrow i}(t)}+\lambda^{S}_{i}(t+1)P^{i}_{S}(t)\prod_{m\in \partial i}\frac{\theta^{m \rightarrow i}(t+1)}{\theta^{m \rightarrow i}(t)}
\\
& + \frac{\epsilon}{\left(\nu_{i}(t) - \underline{\nu_{i}^{t}}\right)}- \frac{\epsilon}{\left(\overline{\nu_{i}^{t}}-\nu_{i}(t)\right)}=0.
\label{eq:variation_mu}
\end{align}

This set of forward and backward equations is sufficient for determining $\{\underline{\nu}_{i}\}_{i \in V}$, using the following optimization procedure, analogous to~\cite{le1988theoretical,saad1997globally}. First, Initialize $\nu_{i}(t)$ to some initial values, for example using the uniform assignment $\nu_{i}(t)=1/NT$ for all $i \in V$ and $t \in [0,T-1]$. Then repeat the following steps for a fixed number of iterations or until convergence:
\begin{enumerate}
\item Starting from the initial values for the dynamics variables~\eqref{app:eq:SIequations:initial_conditions}, propagate the DMP equations~\eqref{app:eq:SIequations:S}-\eqref{app:eq:SImarginals:I} forward, storing the values of dynamic messages at different times.
\item Use equations~\eqref{eq:variation_phi}-\eqref{eq:variation_mu} for fixing the boundary values of Lagrange multipliers at time $T$:
\begin{itemize}
\item~\eqref{eq:variation_phi} assigns $\lambda^{\phi}_{k \rightarrow i}(T)=0$;
\item~\eqref{eq:variation_P_S} gives $\lambda^{S}_{k \rightarrow i}(T)=0$;
\item~\eqref{eq:variation_P_S_marginal} gives $\lambda^{S}_{i}(T)=\delta_{t_{i},T}$;
\item~\eqref{eq:variation_mu} sets $\lambda^{\nu}_{B}(T-1)$ and $\nu_{i}(T-1)$ through $\lambda^{S}_{i}(T)$ and $\lambda^{S}_{k \rightarrow i}(T)$. Writing~\eqref{eq:variation_mu} as
\begin{equation}
\lambda^{\nu}_{B}(t) + \psi_{i}(t) + \frac{\epsilon}{\left(\nu_{i}(t) - \underline{\nu_{i}^{t}}\right)}- \frac{\epsilon}{\left(\overline{\nu_{i}^{t}}-\nu_{i}(t)\right)} = 0,
\end{equation}
we first express $\nu_{i}(T-1)$ for each $i \in V$ as a function of $\lambda^{\nu}_{B}(T-1)$ as a solution of the second order equation. When control parameters can take all possible values between 0 and 1, meaning that $\underline{\nu_{i}^{t}} = 0$ and $\overline{\nu_{i}^{t}} = 1$ for all $i \in V$ and $t \in [0,T-1]$, then there are two a priori possible solutions:
\begin{equation}
\nu^{\pm}_{i}(t) = \frac{\lambda^{\nu}_{B}(t)+\psi_{i}(t) - 2\epsilon \pm \sqrt{(\lambda^{\nu}_{B}(t)+\psi_{i}(t))^{2}+4\epsilon^{2}}}{2\lambda^{\nu}_{B}(t)+2\psi_{i}(t)}.
\label{eq:nu_via_eta}
\end{equation}
Assuming that $\epsilon$ is positive, we choose the solution with the positive sign in front of the square root which always leads to $0 < \nu^{+}_{i}(t) < 1$, thus expressing $\{\nu_{i}(T-1)\}_{i \in V}$ as a function of $\lambda^{\nu}_{B}(T-1)$. We then determine $\lambda^{\nu}_{B}(T-1)$ (and hence $\{\nu^{\pm}_{i}(T-1)\}_{i \in V}$) from the numerical solution of the budget constraint equation~\eqref{app:eq:Budget_nu}.
\item~\eqref{eq:variation_theta} gives $\lambda^{\theta}_{k \rightarrow i}(T)$ through $\lambda^{S}_{i}(T)$, $\lambda^{S}_{k \rightarrow i}(T)$ and $\nu_{i}(T-1)$.
\end{itemize}
\item Using the computed values for dynamic messages at different times, propagate the equations~\eqref{eq:variation_phi}-\eqref{eq:variation_mu} backward to compute the values of the Lagrange multipliers and parameters at all times (compute values at time $t$ given values at times $t+1$):
\begin{itemize}
\item Eq.~\eqref{eq:variation_phi} gives $\lambda^{\phi}_{k \rightarrow i}(t)$;
\item solve~\eqref{eq:variation_P_S} and~\eqref{eq:variation_P_S_marginal} for evaluating $\lambda^{S}_{k \rightarrow i}(t)$ and $\lambda^{S}_{i}(t)$, correspondingly.
\item compute $\lambda^{\nu}_{B}(t-1)$ and $\nu_{i}(t-1)$ using~\eqref{eq:variation_mu} and~\eqref{app:eq:Budget_nu} in the same way as it has been done for $t=T-1$, see equation~\eqref{eq:nu_via_eta};
\item Eq.~\eqref{eq:variation_theta} gives $\lambda^{\theta}_{k \rightarrow i}(t)$ through $\lambda^{\theta}_{k \rightarrow i}(t+1)$, $\lambda^{S}_{i}(t+1)$, $\lambda^{S}_{k \rightarrow i}(t+1)$, $\lambda^{S}_{i}(t)$, $\lambda^{S}_{k \rightarrow i}(t)$, $\nu_{i}(t)$ and $\nu_{i}(t-1)$.
\end{itemize}
\item Using the updated values of $\nu_{i}(t)$ in the step 1.
\end{enumerate}

At each iteration, this procedure leads to ``jumps'' between local optima of the objective $\mathcal{O}$ verifying the budget constraint~\eqref{app:eq:Budget_nu}, where we keep track of the best local optimum which provides the solution to the optimization problem after a maximum number of iterations is reached. Additionally, several different initializations for $\{\nu_{i}(t)\}_{i,t}$ and $\epsilon$ can be used in this scheme in order to achieve the best solution. Note that the initial distribution of $\{\nu_{i}(t)\}_{i,t}$ does not have to satisfy the budget condition~\eqref{app:eq:Budget_nu}, but the solutions obey the constraint already after the first forward-backward iteration. In the case of a small number of nodes in the network $N$ which corresponds to a simpler optimization landscape, we observed that the scheme quickly converges to a unique optimum starting from an arbitrary initial condition. In the next subsection, we describe tests of this scheme on a number of real-world networks.


\subsection{Tests on real-world networks and comparisons with popular heuristics}
\label{sec:Seeding_tests}

For validating the optimization scheme and its performance against existing approaches, we test the algorithm on a particular case of the targeting setting: the optimal seeding problem which corresponds to optimizing the initial conditions. Indeed, the initial conditions at time one can be defined as an optimization over $\{\nu_{i}(0)\}_{i \in V}$ at time $t=0$ at which nodes will have a spontaneous probability of switching to the infected state at time zero. This leads to a more general space of possible initial conditions, without restrictions to a selection of an integer number of infected seeds. A classical setting corresponding to the selection of a group of ``influential spreaders'' at initial time $t=1$ can be recovered in our framework by imposing additional constraints to the domain of variation of $\{\nu_{i}(0)\}_{i \in V}$, restricting them to take only values close to zero or one (e.g. using an appropriate barrier potential).

\subsubsection{Comparison with an explicit symbolic resolution of the DMP equations}

\begin{figure}[!ht]
\begin{center}
\includegraphics[width=0.34\columnwidth]{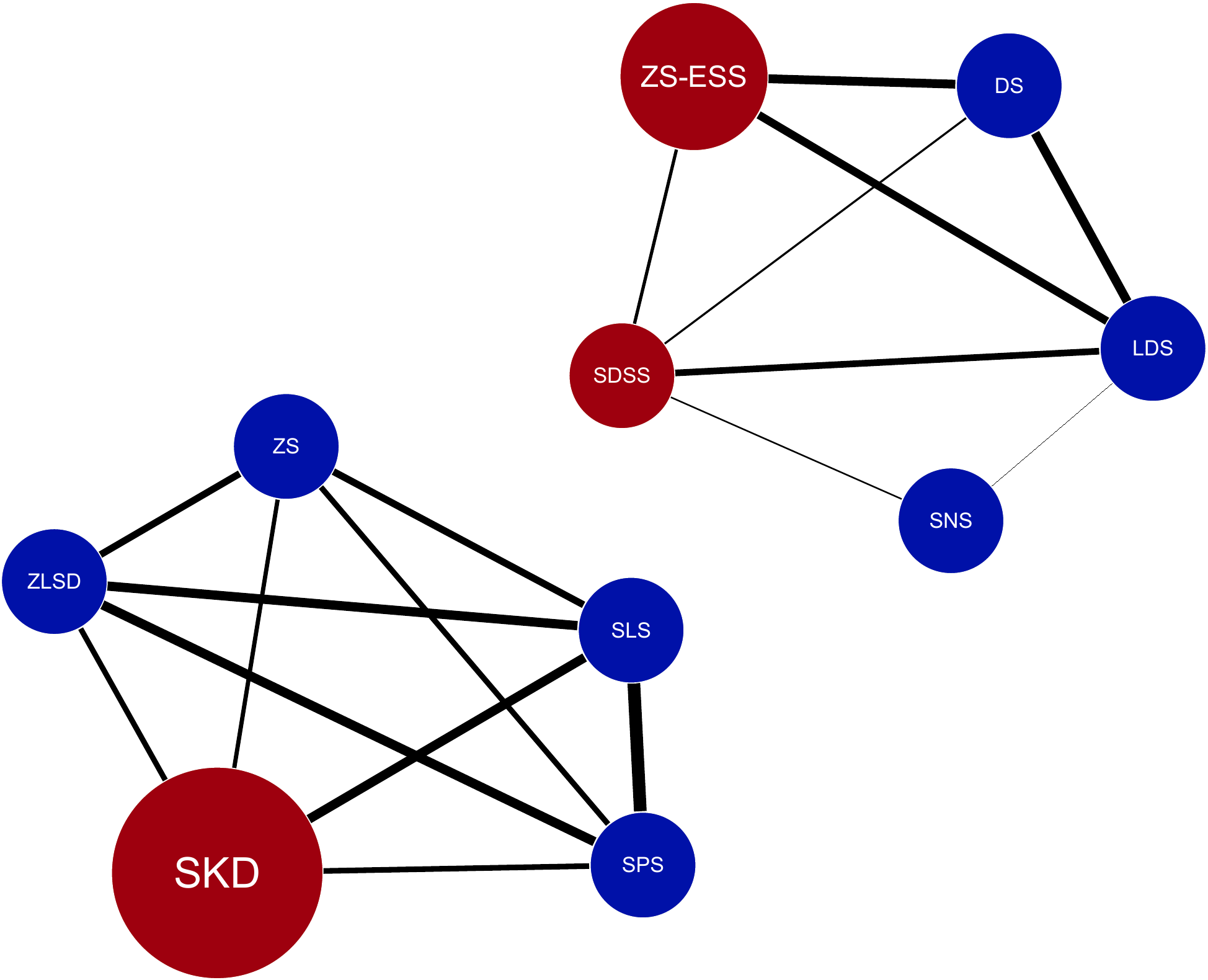}
\caption{A small network of relations between Slovene parliamentary political parties from the social study of~\cite{doreian1996partitioning}. The edge thickness is proportional to the estimated mutual ``influence'' $\alpha_{ij}$ between a pair of parties. Only red nodes are assumed to be controllable in this network. The sizes of the nodes reflect the optimal distribution of the seeding budget $B_{\nu}(0)=1.5$ for maximizing the global impact: small-size nodes do not require any control, and the spontaneous activation probabilities associated with the medium and large-size nodes are equal to $0.5$ and $1.0$, respectively.}
\label{fig:Parties}
\end{center}
\end{figure}

As a validation example considered in the main text, we run the optimization scheme on a small network of relations between Slovene parliamentary parties in 1994, depicted in the Figure~\ref{fig:Parties}. The nodes in this network correspond to Slovene political parties in 1994, and the links represent the similarity relations estimated from a sociological survey of the parliament members which were asked to estimate the distance between pairs of parties in the political space~\cite{doreian1996partitioning}. We have renormalized the weights of edges in such a way that the maximum pairwise mutual ``influence'' receives the value $\alpha^{\max}_{ij} = 0.5$, and the other weights are distributed proportionally to the survey data. We use this real-world network for the first tests in the seeding problem, which was defined as follows. We assumed that three nodes in this network belong to a controllable set $W = {1,3,5}$, corresponding to the parties SKD, ZS-ESS and SDSS, correspondingly. Given the total campaigning budget $B_{\nu}(0) = 1.5$, the goal is to maximize the total informational ``influence'' at time $T=3$ by finding the optimal initial distribution of the associated control parameters $\nu_{1}(0)$, $\nu_{5}(0)$ and $\nu_{3}(0) = 1.5 - \nu_{1}(0) - \nu_{5}(0)$.

As suggested in subsection~\ref{sec:Seeding_tests}, the forward-backward optimization scheme in this case converges to a unique optimal solution in about $7$ iterations with $\epsilon = 5 \cdot 10^{-4}$ starting from an arbitrary initial values of the control parameters, outputting the values $\nu^{*}_{1}(0) = 0.993217$ and $\nu^{*}_{5}(0) = 0.505758$. One advantage of this small test case is that we can check of the validity of the scheme via an explicit symbolic resolution of DMP equations, obtaining a closed-form expression of the objective function $\mathcal{O} = \sum_{i \in V} P_{I}^{i}(T)$, representing the final spread at time $T$ as a function of the independent control parameters $\nu_{1}(0)$ and $\nu_{5}(0)$:
\begin{align*}
\mathcal{O} = &-0.0213192 \nu_{5}(0)^{3} \nu_{1}(0)-0.0113728 \nu_{5}(0)^{4}-0.00149759 \nu_{5}(0)^{3}+0.125857 \nu_{5}(0)^{2} \nu_{1}(0)
\\
&+1.19019 \nu_{5}(0)^{2}
+ 1.05873 \nu_{5}(0) \nu_{1}(0)-1.5701 \nu_{5}(0)-0.00130489 \nu_{1}(0)^{4}+0.0456274 \nu_{1}(0)^{3}
\\
&
-0.754842 \nu_{1}(0)^{2}+1.74962 \nu_{1}(0)
-0.00124248 \nu_{1}(0)^{3} \nu_{5}(0)^{2}
- 0.00274613 \nu_{1}(0)^{2} \nu_{5}(0)^{3}
\\
&
-0.00470875 \nu_{1}(0)^{2} \nu_{5}(0)^{2}+0.00523158 \nu_{1}(0)^{3} \nu_{5}(0)
+0.121085 \nu_{1}(0)^{2} \nu_{5}(0)-0.0017293 \nu_{1}(0) \nu_{5}(0)^{4}
\\
&
- 2.9969\cdot 10^{-6} \nu_{1}(0)^{4} \nu_{5}(0)^{2}+7.3384\cdot 10^{-5} \nu_{1}(0)^{3} \nu_{5}(0)^{3}
+8.88018\cdot 10^{-6} \nu_{1}(0)^{4} \nu_{5}(0)
\\
&
+0.000238133 \nu_{1}(0)^{2} \nu_{5}(0)^{4}+ 0.000244127 \nu_{1}(0) \nu_{5}(0)^{5}
-0.000234529 \nu_{5}(0)^{5}+8.23747\cdot 10^{-5} \nu_{5}(0)^{6}
\\
&
+2.10571\cdot 10^{-7} \nu_{5}(0)^{3} \nu_{1}(0)^{4}+8.42285\cdot 10^{-7} \nu_{5}(0)^{4} \nu_{1}(0)^{3}
+ 1.26343\cdot 10^{-6} \nu_{5}(0)^{5} \nu_{1}(0)^{2}
\\
&
+8.42285\cdot 10^{-7} \nu_{5}(0)^{6} \nu_{1}(0)+2.10571\cdot 10^{-7} \nu_{5}(0)^{7}+3.63985.
\end{align*}
The optimal parameter values can be obtained by a direct maximization of $\mathcal{O}(\nu_{1}(0),\nu_{5}(0))$ plotted in Figure~$2$ of the main text. The optimal parameter values are given in this case by $\nu^{*}_{1}(0) = 1.0$ and $\nu^{*}_{5}(0) = 0.5$, see Figure~\ref{fig:Parties}, in full agreement with the solution of the forward-backward iteration scheme up to domain border perturbations due to the finite value of $\epsilon$.

\subsubsection{Description of popular heuristics used for comparisons on large real-world networks}

In the main text, we described the performance of the DMP algorithm for the seeding problem on real-world networks of different nature, topology and size, and presented comparisons with well-preforming heuristics in the case of deterministic spreading in the SI model. Here we briefly describe the algorithms used for comparisons.

\begin{enumerate}
\item {\bf Naive strategies}. As a reference, we present results obtained by a naive allocation of budget to randomly-selected nodes ({\bf Random}) and a uniform allocation of the budget in a probabilistic way ({\bf Uniform}).
\item {\bf High degree adaptive (HDA)}. The algorithm is based on the idea that the best spreaders correspond to nodes with the highest degree~\cite{pastor2002immunization}, which is a valid and reasonable assumption for a large number of models, including the SI model considered here. The budget is iteratively attributed to the nodes with the current largest degree. Note the adaptive nature of the algorithm: once the node is selected, it is effectively removed from the network, and the degrees of all nodes are recomputed accordingly; this strategy is much more efficient than the allocation of the budget to the high-degree nodes computed once and for all nodes in the original graph.
\item {\bf $k$-shell decomposition}. In this algorithm, the influential spreaders are ranked according to their belonging to the $k$-core with maximum $k$. However, it has been noted in~\cite{kitsak2010identification} that while this strategy is successful in identifying a single ``influential'' node, it performs badly when a group of nodes is selected, which is confirmed by our findings.
\item {\bf Collective influence}. The relation between the destruction of the giant component and ``optimal percolation'' has been suggested in~\cite{morone2015influence}, where the authors have put forward another topological centrality measure called Collective Influence, defined as 
\begin{equation}
\text{CI}_{l} = (d_{i}-1)\sum_{j \in \partial B(i,l)}(d_{j}-1),
\end{equation}
where $\partial B(i,l)$ denotes the set of nodes at a distance $l$ from node $i$. This topological characteristic results from mapping the spreading process asymptotically onto percolation~\cite{newman2002spread} (in a particular instance of the Linear Threshold model~\cite{granovetter1978threshold}). The intuition behind this measure is that the ``collective influence'' of node $i$ is not only given by its degree, but also by the contribution of the degrees of nodes at a certain distance from it. However, on many graphs the number of nodes in the set $\partial B(i,l)$ grows exponentially with $l$, which makes the computation of $\text{CI}_{l}$ rather involved even for small values of $l$. Similarly to HDA, the algorithm is adaptive: after an allocation of resources to the node with the highest score, the measure is recomputed for all nodes. It is intuitive that in the case of a finite-time horizon objective, the best performance should be attained for $l \leq T$; indeed, we find that $l\!=\!2$ realizes the best choice for the algorithm, with CI$_{4}$ leading to sub-optimal results in this near-deterministic spreading case.
\end{enumerate}

Note that studying the minimization of the spread on networks is less interesting, because the optimal solution corresponds to activating nodes in clusters which are disconnected from the majority of nodes belonging to the giant component.

\section{Offline and online mitigation of epidemic via optimal vaccination}
\label{app:C}

Here we consider the problem of spread minimization, for example in the case of an undesired spreading process such as the propagation of epidemic. The minimization of the spread at time $T$ is achieved by an effective distribution of vaccines to vulnerable susceptible nodes in a dynamical fashion. We use the variant of the generalized SIR model with a vaccination transition: the optimization parameters in this case are given by $\{\mu_{i}(t)\}_{i \in V}$ for $t \in [0,T-1]$, and we assume that the spontaneous transition to the $I$ state is absent, meaning that $\nu_{i}(t)=0$ for all $i$ and $t$. Modifications with respect to the case considered previously in the Section~\ref{Sec:Spread_maximization} include:
\begin{enumerate}
\item We are interested in maximizing the objective $\sum_{i \in V}\left(1-P^{i}_{I}(T)\right) = \sum_{i \in V} \left( P^{i}_{S}(T) + P^{i}_{R}(T) \right)$ at this time, so the objective has a different sign with respect to the previous case of the spread maximization;
\item Since now nodes can transit to the $R$ state, we additionally need to keep track of the evolution of the quantities $P^{i}_{R}(t)$ and $P^{i \rightarrow j}_{R}(t)$;
\item Similarly to the formulation in the spread maximization, it is possible to include additional information on a list of target nodes which require priority protection, thus forcing the optimization to minimize the probability of infection on ``key'' nodes first;
\item In the setting of ``online'' mitigation of the spreading process one should reinitialize the optimization process at each time step once the new state of the network is available, getting a more accurate and updated account of the state of the system. One then solves the optimization problem of spread minimization at the horizon $t=T$ for the allocation of resources at the next time step only (seeding problem) using new data as initial conditions for the DMP equations. Running optimization for a few more time steps results in a vaccine distribution plan with forcasted estimates, which might be an important feature in realistic settings.
\end{enumerate}

\subsection{Lagrangian and forward-backward equations}

The corresponding Lagrangian in the case of spreading minimization under vaccination constraints takes the following form:

{\small
\begin{align}
\notag
\mathcal{L}&=\underbrace{\sum_{i\in V}\left(P^{i}_{S}(T)+P^{i}_{R}(T)\right)}_{\mathcal{O}}
+\underbrace{\sum_{t=0}^{T-1} \lambda^{\mu}_{B}(t) \left[\sum_{i \in V} \mu_{i}(t) - B_{\mu}(t) \right]}_{\mathcal{B}}
+\underbrace{\epsilon \sum_{t=0}^{T-1} \sum_{i \in V} \left( \log \left[ \mu_{i}(t)- \underline{\mu_{i}^{t}} \right] + \log \left[\overline{\mu_{i}^{t}} - \mu_{i}(t) \right] \right)}_{\mathcal{P}}
\\
\notag
&\left.
\begin{aligned}
&+\sum_{i\in V}\sum_{t=0}^{T-1}\lambda^{S}_{i}(t+1)\left[P_{S}^{i}(t+1)-P_{S}^{i}(t)(1-\mu_{i}(t))\prod_{k \in \partial i}\frac{\theta^{k \rightarrow i}(t+1)}{\theta^{k \rightarrow i}(t)}\right]
\\
&+\sum_{i\in V}\sum_{t=0}^{T-1}\lambda^{R}_{i}(t+1)\left[P_{R}^{i}(t+1)-P_{R}^{i}(t) -\mu_{i}(t)P_{S}^{i}(t)\right]
\\
&+\sum_{(ki)\in E}\sum_{t=0}^{T-1}\lambda^{S}_{k \rightarrow i}(t+1)\left[P_{S}^{k \rightarrow i}(t+1)-P_{S}^{k \rightarrow i}(t)(1-\mu_{k}(t))\prod_{l\in \partial k \backslash i}\frac{\theta^{l \rightarrow k}(t+1)}{\theta^{l \rightarrow k}(t)}\right]
\\
&+\sum_{(ki)\in E}\sum_{t=0}^{T-1}\lambda^{R}_{k \rightarrow i}(t+1)\left[P_{R}^{k \rightarrow i}(t+1)-P_{R}^{k \rightarrow i}(t)-\mu_{k}(t)P_{S}^{k \rightarrow i}(t)\right]
\\
&+\sum_{(ki)\in E}\sum_{t=0}^{T-1}\lambda^{\theta}_{k \rightarrow i}(t+1)\left[\theta^{k \rightarrow i}(t+1)-\theta^{k \rightarrow i}(t)
+\alpha_{ki}\phi^{k \rightarrow i}(t)\right]
\\
&+\sum_{(ki)\in E}\sum_{t=0}^{T\!-\!1}\lambda^{\phi}_{k \rightarrow i}(t\!+\!1)\Big[\phi^{k \rightarrow i}(t\!+\!1)\!-\!(1\!-\!\alpha_{ki})\phi^{k \rightarrow i}(t)
\!-\!P_{S}^{k \rightarrow i}(t)\!+\!P_{S}^{k \rightarrow i}(t\!+\!1)\!-\!P_{R}^{k \rightarrow i}(t)\!+\!P_{R}^{k \rightarrow i}(t\!+\!1)\Big]
\end{aligned}
\right\rbrace\mathcal{D}
\\
\notag
&\left.
\begin{aligned}
&+\sum_{i\in V}\lambda^{S}_{i}(0)\left[P^{i}_{S}(0)-\delta_{\sigma_{i}^{0},S}\right]+\sum_{i\in V}\lambda^{R}_{i}(0)\left[P^{i}_{R}(0)-\delta_{\sigma_{i}^{0},R}\right]+\sum_{(ki)\in E}\lambda^{S}_{k \rightarrow i}(0)\left[P_{S}^{k \rightarrow i}(0)-\delta_{\sigma_{k}^{0},S}\right]
\\
&+\sum_{(ki)\in E}\lambda^{R}_{k \rightarrow i}(0)\left[P_{R}^{k \rightarrow i}(0)-\delta_{\sigma_{k}^{0},R}\right]+\sum_{(ki)\in E}\lambda^{\theta}_{k \rightarrow i}(0)\left[\theta^{k \rightarrow i}(0)-1\right]+\sum_{(ki)\in E}\lambda^{\phi}_{k \rightarrow i}(0)\left[\phi^{k \rightarrow i}(0)-\delta_{\sigma_{k}^{0},I}\right].
\end{aligned}
\right\rbrace\mathcal{I}
\end{align}
}

The variation of the Lagrangian with respect to the dual variables yield the forward DMP equations~\eqref{app:eq:SIequations:S}-\eqref{app:eq:SImarginals:I}. The derivation of the backward equations follow the same principles as in the case of the SI model, with obvious modifications: change of the control parameters $\nu_{i}(t) \rightarrow \mu_{i}(t)$, additional equations
\begin{align}
&\partial \mathcal{L} / \partial P_{R}^{i}(t) = \mathds{1}[t=T] + \lambda_{i}^{R}(t) - \lambda_{i}^{R}(t+1)\mathds{1}[t\neq T] = 0,
\\
&\partial \mathcal{L} / \partial P_{R}^{k \rightarrow i}(t) = \lambda_{k \rightarrow i}^{R}(t) - \lambda_{k \rightarrow i}^{R}(t+1)\mathds{1}[t\neq T] + \lambda^{\phi}_{k \rightarrow i}(t) - \lambda^{\phi}_{k \rightarrow i}(t+1)\mathds{1}[t\neq T] = 0,
\end{align}
and additions to the previous backward equations due to the new terms
\begin{align}
&\Delta \partial \mathcal{L} / \partial P_{S}^{i}(t) = - \lambda_{i}^{R}(t+1)\mu_{i}(t)\mathds{1}[t\neq T],
\\
&\Delta \partial \mathcal{L} / \partial P_{S}^{k \rightarrow i}(t) = - \lambda_{k \rightarrow i}^{R}(t+1)\mu_{k}(t)\mathds{1}[t\neq T],
\\
&\Delta \partial \mathcal{L} / \partial \mu_{i}(t) = - \lambda_{i}^{R}(t+1)P_{S}^{i}(t) - \sum_{k \in \partial i} \lambda_{i \rightarrow k}^{R}(t+1)P_{S}^{i \rightarrow k}(t).
\end{align}

\subsection{Description of the data set used for tests}

As a test example for the vaccination problem, we constructed a transportation network of busiest flight routes between major U.S.~airports, extracted from the publicly available Bureau of Transportation Statistics data \cite{bts}. We used the data table providing the number of transported passengers by different companies between U.S.~airports over the past several years. First of all, we extracted the sub-table of flights between $61$ biggest airports in terms of the total number of emplaned passengers per year, including $30$ ``major'' and $31$ ``largest'' hubs according to the BTS classification. Then multiple entries corresponding to the same route have been aggregated, and the routes carrying less then $10\%$ of the passengers transported by the busiest route have been pruned as less significant ones, primarily for rending the network reasonably sparse for a clear visualization. This resulted in a network with $M = \vert E \vert\!=\!383$ edges. Finally, we have assigned the spreading couplings ${\alpha_{ij}}_{(ij) \in E}$ proportionally to the number of carried passengers in such a way that the lightest route had the value $\alpha^{\min}_{ij} = 0.05$, and hence the busiest route received the value $\alpha^{\max}_{ij} \simeq 0.495$. This choice is based on a reasonable assumption that the probability of infection transmission along each link is proportional to the number of carried passengers on this route.

\section{Case of continuous dynamics}
\label{sec:Continuous}

The optimization procedure described in this work can be directly applied to the case of continuous dynamics, which might be more relevant in other applications. In the continuous case, the backward equations are obtained through the variation of the Lagrangian resulting in the continuous Euler-Lagrange equations. In this section, we illustrate the approach in the setting of maximizing the spread in the SI model using the continuous version of the DMP equations.

\subsection{Maximization of spread at the time horizon $T$}
\label{sec:MaxSpreadTimeHorizon}
In the continuous case, the marginal probability for node $i$ to be in the state $S$ at time $t$ reads
\begin{equation}
P_{S}^{i}(t)=P_{S}^{i}(0) \left( \int_{0}^{t} dt' e^{-\nu_{i}(t')t'} \right) \prod_{k\in \partial i}\theta^{k \rightarrow i}(t).
\label{eq:continous_P_S}
\end{equation}
In the case of constant rates $\alpha_{ij}$, we define the transmission function as $f_{ij}(t) = \alpha_{ij} e^{- \alpha_{ij} t}$. Then the functions $\theta^{i \rightarrow j}(t)$ are computed as follows~\cite{KarrerNewman2010}:
\begin{align}
\notag
\theta^{i \rightarrow j}(t) & = 1 - \int_0^t d \tau \ f_{ij}(\tau) \left[ 1 - P_S^{i} (0) \left( \int_{0}^{t-\tau} dt' e^{-\nu_{i}(t')t'} \right) \prod_{k \in \partial i \backslash j } \theta^{k \rightarrow i}(t - \tau)  \right]
\\
&= e^{- \alpha_{ij} t} +  P_S^{i} (0) \alpha_{ij} e^{- \alpha_{ij} t} \int_0^t d \tau \  e^{ \alpha_{ij} \tau} \left( \int_{0}^{\tau} dt' e^{-\nu_{i}(t')t'} \right) \prod_{k \in \partial i \backslash j} \theta^{k \rightarrow i}( \tau).
\label{eq:continous_integral}
\end{align}
\old{In order to compute the dynamic messages $\theta^{i \rightarrow j}(t)$, we can either integrate the expression~\eqref{eq:continous_integral} numerically, or transform the equation above into an ordinary differential equation by differentiating~\eqref{eq:continous_integral}:}
\new{In order to compute the dynamic messages $\theta^{i \rightarrow j}(t)$, we can either integrate the Eq.~\eqref{eq:continous_integral} numerically, or transform it into an ordinary differential equation by differentiating with respect to $t$:}
\begin{equation}
\dot{\theta}^{i \rightarrow j}(t) =  - \alpha_{ij} \theta^{i \rightarrow j}(t) +  \alpha_{ij} P_S^i (0) \left( \int_{0}^{t} dt' e^{-\nu_{i}(t')t'} \right) \prod_{k \in \partial i \backslash j} \theta^{k \rightarrow i}(t),
\label{eq:continous_differential}
\end{equation}
\old{which can be solved numerically starting from initial conditions $\theta^{i \rightarrow j}(0)=1$. This equation represents the dynamics of $\theta^{i \rightarrow j}(t)$ (the dot notation represents $\frac{d}{dt}$).}
\new{which represents the dynamics of $\theta^{i \rightarrow j}(t)$ (the dot notation represents $\frac{d}{dt}$) and can be solved numerically starting from initial conditions $\theta^{i \rightarrow j}(0)=1$.}
To derive the dynamics of $P_S^{i} (t)$ one can differentiate~\eqref{eq:continous_P_S} to obtain:
\begin{equation}
\dot{P}_{S}^{i}(t)=P_{S}^{i}(0) e^{-\nu_{i}(t)t} \prod_{k\in \partial i}\theta^{k \rightarrow i}(t) + P_{S}^{i}(t) \sum_{k\in \partial i} \frac{\dot{\theta}^{k \rightarrow i}(t)}{\theta^{k \rightarrow i}(t)}.
\label{eq:continous_P_S_dynamics}
\end{equation}
Maximizing $\mathcal{O} = \sum_{i \in V} P^{i}_{I}(T)$ is equivalent to minimizing
\begin{equation}
\hat{\mathcal{O}}(T)=\sum_{i\in V}P^{i}_{S}(T)=\int_{0}^{T} \sum_{i\in V}\dot{P}^{i}_{S}(t)dt
\label{eq:Cost_Function_continuous}
\end{equation}
under the constraints imposed by~\eqref{eq:continous_differential},~\eqref{eq:continous_P_S_dynamics} and the budget constraint~\eqref{app:eq:Budget_nu}
\begin{equation}
\sum_{i\in V}\nu_{i}(t)=  B_{\nu}(t).
\end{equation}

All constraints will be imposed by the corresponding Lagrange multipliers. The Lagrangian to be extremized takes the form

\begin{eqnarray}
\mathcal{L}&=&\int_{0}^{T} \Biggl\{ \sum_{i\in V}\dot{P}^{i}_{S}(t) + \lambda^{\nu}_{B}(t) \left[\sum_{i\in V}\nu_{i}(t)-B_{\nu}(t)\right] \nonumber   \\
&+& \sum_{(i,j) \in E} \lambda^{\theta}_{i \rightarrow j} (t) \left[ \dot{\theta}^{i \rightarrow j}(t) + \alpha_{ij} \theta^{i \rightarrow j}(t) -  \alpha_{ij} P_S^i (0) \left( \int_{0}^{t} dt' e^{-\nu_{i}(t')t'} \right) \prod_{k \in \partial i \backslash j} \theta^{k \rightarrow i}(t) \right]\nonumber \\
&+& \sum_{i\in V} \lambda^{S}_{i}(t) \left[\dot{P}_{S}^{i}(t)-P_{S}^{i}(0) e^{-\nu_{i}(t)t} \prod_{k\in \partial i}\theta^{k \rightarrow i}(t) - P_{S}^{i}(t) \sum_{k\in \partial i} \frac{\dot{\theta}^{k \rightarrow i}(t)}{\theta^{k \rightarrow i}(t)}  \right]\Biggr\} ~dt \nonumber \\
&=&\int_{0}^{T} \Biggl\{ \sum_{i\in V}\dot{P}^{i}_{S}(t) + \lambda^{\nu}_{B}(t) \left[\sum_{i\in V}\nu_{i}(t)-B_{\nu}(t)\right]  \nonumber \\
&+& \sum_{(i,j) \in E} \lambda^{\theta}_{i \rightarrow j} (t) \left[ \dot{\theta}^{i \rightarrow j}(t) + \alpha_{ij} \theta^{i \rightarrow j}(t) -  \alpha_{ij} \frac{P_S^i (t)}{\theta^{j \rightarrow i}(t)} \right]\nonumber \\
&+& \sum_{i\in V} \lambda^{S}_{i}(t) \left[\dot{P}_{S}^{i}(t)-P_{S}^{i}(0) e^{-\nu_{i}(t)t} \prod_{k\in \partial i}\theta^{k \rightarrow i}(t) - P_{S}^{i}(t) \sum_{k\in \partial i} \frac{\dot{\theta}^{k \rightarrow i}(t)}{\theta^{k \rightarrow i}(t)}  \right]\Biggr\} ~dt
\label{eq:Lagrangian_continuous}
\end{eqnarray}

Variational maximization of a Lagrangian
 \[ \mathcal{L}=\int_{0}^{T} F(t,y,\dot{y}) ~dt \]
  where $F(t,y,\dot{y})$ is a function of some variable $t$, $y$ and $\dot{y}$ its derivative with respect to $t$, is carried out by varying $y$ with respect to a small perturbation $\epsilon\rightarrow 0$, throughout the $t$ interval~\cite{Hildebrand1965}, to $y+\epsilon \delta y$. Extremizing $\mathcal{L}$ by setting $\frac{d \mathcal{L}}{d\epsilon}=0$ and using integration by parts one obtains the optimization condition
  \[ \frac{d \mathcal{L}}{d\epsilon}\bigg|_{\epsilon=0}= \int_{0}^{T} \left[\frac{\partial F}{\partial y} - \frac{d}{dt} \frac{\partial F}{\partial \dot{y}} \right] \delta y~ dt + \left[\frac{\partial F}{\partial \dot{y}} \right]^{T}_{0} =0 ~.\]
  Both the Euler-Lagrange equation $\left[\frac{\partial F}{\partial y} - \frac{d}{dt} \frac{\partial F}{\partial \dot{y}}\right]$ and the boundary conditions  $\left[\frac{\partial F}{\partial \dot{y}} \right]^{T}_{0}$ should be zero.

Varying~\eqref{eq:Lagrangian_continuous} with respect to $\delta P_{S}^{i}(t)$ and $\delta\theta^{i \rightarrow j}(t)$
results in the following differential equations:
\begin{eqnarray}
\dot{\lambda}^{S}_{i}(t) &=& - \sum_{j\in\partial i} \frac{\lambda^{\theta}_{i \rightarrow j}\alpha_{ij}}{\theta^{j \rightarrow i}(t)} - \lambda^{S}_i (t) \sum_{k\in \partial i} \frac{\dot{\theta}^{k \rightarrow i}(t)}{\theta^{k \rightarrow i}(t)} - \delta_{t,0} \lambda^{S}_i (0) \prod_{k \in \partial i } \theta^{k \rightarrow i}(0), \label{eq:Lagrangemultipliers_nu} \\
\dot{\lambda}^{\theta}_{i \rightarrow j}(t) &=& \lambda^{\theta}_{i \rightarrow j}(t)\alpha_{ij} + \lambda^{\theta}_{j \rightarrow i}(t)\alpha_{ji} \frac{P_{S}^{j}(t)}{\left[\theta^{i \rightarrow j}(t)\right]^2} \label{eq:Lagrangemultipliers_lambda}\\
&-& \lambda^{S}_{j}(t) \left\{ P_{S}^{j}(0) e^{-\nu_{j}(t)t} \prod_{k\in \partial j / i}\theta^{k \rightarrow j}(t) - \frac{\dot{P}_{S}^{j}(t)}{\theta^{i \rightarrow j}(t)}\right\} + \dot{\lambda}^{S}_{j}(t)\frac{P_{S}^{j}(t)}{\theta^{i \rightarrow j}(t)}~, \nonumber \\
 &=& \lambda^{\theta}_{i \rightarrow j}(t)\alpha_{ij} - \sum_{k\in \partial j / i} \frac{\lambda^{\theta}_{j \rightarrow k}(t)\alpha_{jk}P_{S}^{j}(t)}{\theta^{k \rightarrow j}(t)\theta^{i \rightarrow j}(t)} - \delta_{t,0} \lambda^{S}_j (0) P_{S}^{j}(t) \prod_{k \in \partial j /i} \theta^{k \rightarrow i}(0),\nonumber
\end{eqnarray}
where the expression for $\dot{\theta}^{k \rightarrow i}(t)$ in both equations~\eqref{eq:Lagrangemultipliers_nu} and~\eqref{eq:Lagrangemultipliers_lambda} are calculated on the basis of~\eqref{eq:continous_differential} and the expressions for $\dot{\lambda}^{S}_{j}(t)$ and $\dot{P}_{S}^{j}(t)$ in~\eqref{eq:Lagrangemultipliers_lambda} are taken from~\eqref{eq:Lagrangemultipliers_nu} and~\eqref{eq:continous_P_S_dynamics}, respectively, and were employed in the simplifications of~\eqref{eq:Lagrangemultipliers_lambda}. These equations will be integrated back on the basis of end points.

Varying~\eqref{eq:Lagrangian_continuous} with respect to $\delta \nu_i (t)$ one obtains:
\begin{equation}
\lambda^{\nu}_{B}(t) = -t \lambda^{S}_{i}(t) P_{S}^{i}(0) e^{-\nu_{i}(t)t} \prod_{k\in \partial i}\theta^{k \rightarrow i}(t)~,
\label{eq:vary_mu}
\end{equation}
which can be re-written as
\begin{equation}
\nu_{i}(t) = -\frac{1}{t} \ln \left[-\frac{\lambda^{\nu}_{B}(t)}{t \lambda^{S}_{i}(t) P_{S}^{i}(0)  \prod_{k\in \partial i}\theta^{k \rightarrow i}(t)}\right]~.
\label{eq:muFunceta}
\end{equation}
Using the condition $\sum_{i\in V} \nu_{i}(t) = B_{\nu}(t)$ we obtain
\begin{equation}
B_{\nu}(t) = -\frac{N}{t} \ln \lambda^{\nu}_{B}(t)+ \frac{1}{t} \sum_{i\in V} \ln \left[-t \lambda^{S}_{i}(t) P_{S}^{i}(0)  \prod_{k\in \partial i}\theta^{k \rightarrow i}(t)\right]~.
\label{eq:muCondition}
\end{equation}
That leads to a straightforward solution for $\lambda^{\nu}_{B}(t)$
\begin{equation}
\lambda^{\nu}_{B}(t) = \exp \Biggl\{ \frac{1}{N} \sum_{i\in V} \ln \left[-t \lambda^{S}_{i}(t) P_{S}^{i}(0)  \prod_{k\in \partial i}\theta^{k \rightarrow i}(t)\right] - \frac{t}{N} B_{\nu}(t) \Biggr\} ~,
\label{eq:etaSolution}
\end{equation}
from which the $\nu_{i}(t)$ values can be calculated.

\textbf{Remark:} An alternative path to the derivation of~\eqref{eq:muFunceta} and~\eqref{eq:etaSolution} consists in directly using the normalized representation \[ \nu_{i}(t)=\frac{e^{-\beta_{i}(t)}}{\sum_{j\in V}e^{-\beta_{j}(t)}} B_{\nu}(t)~.\] instead of enforcing the budget constraint with the Lagrange multiplier $\lambda^{\nu}_{B}(t)$ in~\eqref{eq:Lagrangian_continuous}.
For the optimization with respect to $\beta_{i}(t)$ in this case it is then possible to use
 \[ \frac{\partial}{\partial\beta_k} = -\sum_{j\in V} \left[ \delta_{jk} \nu_k (t) - \nu_k (t)\nu_j (t) \right]  \frac{\partial}{\partial\nu_j}. \]
Optimization with respect to $\beta_{i}(t)$ gives
\begin{equation}
\nu_i (t)= \frac{\lambda^{S}_{i}(t)P_{S}^{i}(0)e^{-\nu_{i}(t)t} \prod_{k\in \partial i}\theta^{k \rightarrow i}(t)}{\sum_{j\in V} \lambda^{S}_{j}(t)P_{S}^{j}(0)e^{-\nu_{j}(t)t} \prod_{k\in \partial j}\theta^{k \rightarrow j}(t)} B_{\nu}(t)~,
\label{eq:Update_mu}
\end{equation}
which is equivalent to~\eqref{eq:muFunceta} and~\eqref{eq:etaSolution}. $\blacksquare$

Additionally, one should enforce the boundary conditions in the two sets of equations:
\begin{eqnarray}
0 &=& \delta P^{i}_{S}(T) \left[ 1 + \lambda^{S}_{i}(T)\right] - \delta P^{i}_{S}(0) \left[ 1 + \lambda^{S}_{i}(0)\right]  \label{eq:Boundaryconditions_nu}\\
0 &=& \delta \theta^{i \rightarrow j}(T) \left[ \lambda^{\theta}_{i \rightarrow j}(T) - \frac{P_{S}^{j}(T)\lambda^{S}_{j}(T) }{\theta^{j \rightarrow i}(T)} \right] - \delta \theta^{i \rightarrow j}(0)\left[ \lambda^{\theta}_{i \rightarrow j}(0) - \frac{P_{S}^{j}(0)\lambda^{S}_{j}(0) }{\theta^{j \rightarrow i}(0)} \right]~.
\label{eq:Boundaryconditions_lambda}
\end{eqnarray}
Since $\delta P^{i}_{S}(0)=\delta \theta^{i \rightarrow j}(0)=0 ~\forall i,j\in V$ \new{it provides the end conditions:}
\begin{eqnarray}
\lambda^{S}_{i}(T) &=&-1   \label{eq:BoundaryconditionsT_nu}\\
 \lambda^{\theta}_{i \rightarrow j}(T) &=& \frac{P_{S}^{j}(T)}{\theta^{j \rightarrow i}(T)} ~.
\label{eq:BoundaryconditionsT_lambda}
\end{eqnarray}
Note that $\lambda^{S}_{i}(T)=-1$ is a result of the minimization of $\hat{\mathcal{O}}(T)$; a maximization of $\mathcal{O}$ would provide a boundary condition of $\lambda^{S}_{i}(T)=1$.

The optimization process should be carried out as follows:
\begin{enumerate}
\item Using the initial conditions $\theta^{i \rightarrow j}(0)=1~\forall i$ and $P^{i}_{S}(0)$ according to the case studied, and some valid initial set of  $\nu_{i}(t)$ (can be uniform), one can solve forward the equations~\eqref{eq:continous_differential} and~\eqref{eq:continous_P_S_dynamics} to obtain (and register) values throughout the dynamics $t=0\rightarrow T$.
\item Using the boundary conditions~\eqref{eq:BoundaryconditionsT_lambda} and~\eqref{eq:BoundaryconditionsT_nu} one solve backward~\eqref{eq:Lagrangemultipliers_nu} and~\eqref{eq:Lagrangemultipliers_lambda} and updates the values of $\nu_{i}(t)$ according to~\eqref{eq:etaSolution} and~\eqref{eq:muFunceta}.
\item The process is repeated until it converges and the final $\nu_{i}(t)$ values represent the solution.
\end{enumerate}

\subsection{Continuous dynamics with targeted and accessible nodes}

In this variant of the problem one targets specific nodes $i \in U$ where $U\subseteq V$ is the subset of all nodes $V$, aiming to maximize the impact at predefined times $t_i$, which may be different for each of the nodes. We will define $T\equiv \max_{i \in U} t_i$. We also assume one has access to a subset on the nodes $i \in W$ where $W\subseteq V$ and $W \cap U = \emptyset$. Again, maximizing $\mathcal{O}$ is equivalent to minimizing
\begin{equation}
\hat{\mathcal{O}}=\sum_{i\in U}P^{i}_{S}(t_i)=\sum_{i\in U}\int_{0}^{t_i}\dot{P}^{i}_{S}(t)~dt~.
\label{eq:Cost_Function_continuous_targeted}
\end{equation}
Since the budget for nodes $i \notin W$ is zero by definition $\nu_{i}(t)=0, \forall i\notin W, \forall t,$ and
the cost constraint corresponding to~\eqref{app:eq:Budget_nu} becomes
\begin{equation}
\label{eq:Constraint on mu_accessible}
\sum_{i\in W}\nu_{i}(t)=  B_{\nu}(t)~,
\end{equation}
where we can use the normalized representation where $\beta_{i}(t)=-\infty, \forall i\notin W, \forall t,$\[ \nu_{i}(t)=\frac{e^{-\beta_{i}(t)}}{\sum_{j\in V}e^{-\beta_{j}(t)}} B_{\nu}(t)~.\]

The Lagrangian to be extremized takes the form (the budget constraint on accessible nodes~\eqref{eq:Cost_Function_continuous_targeted} is embedded in the corresponding $\nu$ variables through $\beta$, as \old{explained in the Remark in the previous section)} \new{ remarked in Sec.~\ref{sec:MaxSpreadTimeHorizon}}
\begin{eqnarray}
\mathcal{L}&=& \sum_{i\in U}\int_{0}^{t_i}\dot{P}^{i}_{S}(t) ~dt \nonumber
\\
&+& \int_{0}^{T}\Biggl\{ \sum_{(i,j) \in E} \lambda^{\theta}_{i \rightarrow j} (t) \left[ \dot{\theta}^{i \rightarrow j}(t) + \alpha_{ij} \theta^{i \rightarrow j}(t) -  \alpha_{ij} P_S^i (0) \left( \int_{0}^{t} dt' e^{-\nu_{i}(t')t'} \right) \prod_{k \in \partial i \backslash j} \theta^{k \rightarrow i}(t) \right]\nonumber \\
&+& \sum_{i\in V} \lambda^{S}_{i}(t) \left[\dot{P}_{S}^{i}(t)-P_{S}^{i}(0) e^{-\nu_{i}(t)t} \prod_{k\in \partial i}\theta^{k \rightarrow i}(t) - P_{S}^{i}(t) \sum_{k\in \partial i} \frac{\dot{\theta}^{k \rightarrow i}(t)}{\theta^{k \rightarrow i}(t)}  \right]\Biggr\} ~dt \nonumber \\
&= & \sum_{i\in U} \int_{0}^{t_i}\dot{P}^{i}_{S}(t) ~dt \nonumber \\
&+& \int_{0}^{T}\Biggl\{\sum_{(i,j) \in E} \lambda^{\theta}_{i \rightarrow j} (t) \left[ \dot{\theta}^{i \rightarrow j}(t) + \alpha_{ij} \theta^{i \rightarrow j}(t) -  \alpha_{ij} \frac{P_S^i (t)}{\theta^{j \rightarrow i}(t)} \right]\nonumber \\
&+& \sum_{i\in V} \lambda^{S}_{i}(t) \left[\dot{P}_{S}^{i}(t)-P_{S}^{i}(0) e^{-\nu_{i}(t)t} \prod_{k\in \partial i}\theta^{k \rightarrow i}(t) - P_{S}^{i}(t) \sum_{k\in \partial i} \frac{\dot{\theta}^{k \rightarrow i}(t)}{\theta^{k \rightarrow i}(t)}  \right]\Biggr\} ~dt
\label{eq:Lagrangian_continuous targeted}
\end{eqnarray}
Alternatively, the first term can be written as \[\int_{0}^{T} \sum_{i\in U}\dot{P}^{i}_{S}(t) \Theta(t_i -t)~dt. \]

Since the first term does not contribute to the Euler-Lagrange equation, varying~\eqref{eq:Lagrangian_continuous targeted} with respect to $\delta P_{S}^{i}(t),~\delta\theta^{i \rightarrow j}(t)$ and  $\delta\nu_{i}(t)$ results in the same differential equations for $\dot{\lambda}^{S}_{i}(t)$~\eqref{eq:Lagrangemultipliers_nu} and $\dot{\lambda}_{ij}(t)$~\eqref{eq:Lagrangemultipliers_lambda}.

Optimization with respect to $\beta_{i}(t)$ or $\nu_i (t)$ gives a similar expression for $i \in W$~\eqref{eq:Update_mu} but $\nu_i (t)=0, \forall i \notin W$ and $\forall t$. Additionally, one should enforce the boundary conditions in the two sets of equations~\eqref{eq:Boundaryconditions_nu} and~\eqref{eq:Boundaryconditions_lambda}. For $j \in U$~\eqref{eq:Boundaryconditions_nu} and~\eqref{eq:Boundaryconditions_lambda} are the same as in the non-targeted case, but for $j \notin U$ there is no constant (\new{of value} 1) in~\eqref{eq:Boundaryconditions_nu} leading to:
\begin{eqnarray}
\lambda^{S}_{i}(T) &=& \begin{cases} -1 & \forall~ i \notin U  \\
0  & \forall~ i \notin U  \end{cases}  \label{eq:BoundaryconditionsT_nu nontargets}\\
 \lambda^{\theta}_{i \rightarrow j}(t_j) &=& \begin{cases} 0 & \forall~ i \notin U \\
 \frac{P_{S}^{j}(t_j)}{\theta^{j \rightarrow i}(t_j)} &\forall~ j \in U  \end{cases}  .
\label{eq:BoundaryconditionsT_lambda nontargets}
\end{eqnarray}
The optimization process should be carried out as before.

\normalsize

\twocolumngrid
\bibliographystyle{naturemag}
\bibliography{model}

\end{document}